\newif\ifsingle

\newif\ifFullVersion
\FullVersiontrue 

\ifsingle
\documentclass[11pt,draftclsnofoot, onecolumn]{IEEEtran}		
\else		
\documentclass[10pt,final, twocolumn]{IEEEtran}
\fi

\usepackage{HKF_pkg}
\usepackage{HKF_ac}
\usepackage{HKF_cmd}

\title{HKF: Hierarchical Kalman Filtering \\
with Online Learned Evolution Priors\\
for Adaptive ECG Denoising 
}
\author{  
\IEEEauthorblockN{Guy Revach, Timur Locher, Nir Shlezinger, Ruud J. G. van Sloun, and Rik Vullings\\ 
} 
\thanks{ 
Parts of this work were presented at the IEEE International Conference on Acoustics, Speech, and Signal Processing (ICASSP) 2023~\cite{locher2022hierarchical}.
 G. Revach and T. Locher and are with the Institute for Signal and Information Processing (ISI), D-ITET, ETH Zürich, Switzerland (e-mail: grevach@ethz.ch). 
N. Shlezinger is with the School of ECE, Ben-Gurion University of the Negev, Beer Sheva, Israel (e-mail: nirshl@bgu.ac.il). 
R. J. G. van Sloun is with the EE Dpt., Eindhoven University of Technology, The Netherlands, (e-mail: r.j.g.v.sloun@tue.nl).
R. Vullings is with the EE Dpt., Eindhoven University of Technology, and with Nemo Healthcare, Veldhoven, The Netherlands, (e-mail: r.vullings@tue.nl).
We thank Hans-Andrea Loeliger for helpful discussions, and Mehdi Bakka for helping with the empirical evaluation.}
}

\begin{document}
\maketitle
\pagestyle{plain}
\thispagestyle{plain}
%
%
\begin{abstract}
\Ac{ecg} signals play a pivotal role in many healthcare applications, especially in at-home monitoring of vital signs. Wearable technologies, which these applications often depend upon, frequently produce low-quality \ac{ecg} signals. While several methods exist for \ac{ecg} denoising to enhance signal quality and aid clinical interpretation, they often underperform with \ac{ecg} data from wearable technology due to limited noise tolerance or inadequate flexibility in capturing \ac{ecg} dynamics. This paper introduces \acs{hkf}, a hierarchical and adaptive Kalman filter, which uses a proprietary state space model to effectively capture both intra- and inter-heartbeat dynamics for \ac{ecg} signal denoising. \acs{hkf} learns a patient-specific structured prior for the \ac{ecg} signal's intra-heartbeat dynamics in an online manner, resulting in a filter that adapts to the specific \ac{ecg} signal characteristics of each patient. In an empirical study, \acs{hkf} demonstrated superior denoising performance (reduced \acl{mse}) while preserving the unique properties of the waveform. In a comparative analysis, \acs{hkf} outperformed previously proposed methods for \ac{ecg} denoising, such as the model-based Kalman filter and data-driven autoencoders. This makes it a suitable candidate for applications in extramural healthcare settings.
\end{abstract}
\acresetall 
%
%
\section{Introduction}\label{sec:Intro}
%
%
In the pursuit to halt the increase in healthcare costs and create a sustainable healthcare system, progressively more patients should be monitored outside the hospital environment. In recent years, various technologies have emerged for remote and ambulatory monitoring of vital signs, including wearable \ac{ecg} monitoring devices~\cite{Bai2021,Sana2020}. 


%
The \ac{ecg} reflects the electrical activity of the heart and is considered one of the most important and informative monitoring modalities, which reveals information about cardiac function and possible pathologies.
%
More specifically, an \ac{ecg} can play a large part in the clinical detection of diseases, including coronary heart diseases, heart attacks, and arrhythmia that can lead to even more severe conditions, such as stroke~\cite{NHS, NHS_Arrhythmia}. \ac{ecg} signal analysis can also play a crucial part in detecting the asphyxia of a fetus during labor~\cite{amerwahlin2001}. 

The specific shape of the \ac{ecg} waveform is used by medical professionals and cardiologists to diagnose the specific heart condition and, therefore, it is essential that the \ac{ecg} recording is as clean as possible. To diagnose the specific condition, or deterioration thereof, of a patient, a medical professional primarily focuses on specific characteristics in the \ac{ecg}. These characteristics can differ between applications. For instance, in case of myocardial infarction or hypoxia in the fetus, the ST-segment often provides vital information~\cite{amerwahlin2001,NHS_Heart_attack}. Monitoring the ST-interval or the occurrence of a negative T-wave amplitude can also be beneficial since these indicate compromised cardiac performance \cite{maclachlan1992fetal, kazmi2011st}.
%
%

Compared with in-hospital monitoring, at-home \ac{ecg} monitoring comes at the expense of signal quality, e.g., electrodes incorporated in garments that are used for recording the \ac{ecg} generally provide {noisier} signals, with more artifacts than the adhesive electrodes that are typically used in the hospital~\cite{Gruetzmann2007}. Although simple filtering can suppress certain noises and artifacts~\cite{Bailey1990}, its effectiveness is limited for \ac{agn}, due to a partial overlap between the signal and noise bandwidth~\cite{velayudhan2016noise}. The challenge of denoising ECG signals corrupted by AGN is the primary focus of this paper.
In the field of \ac{ecg} denoising, recent literature presents a variety of approaches, ranging from classical signal processing techniques to cutting-edge \acl{dl} methods. Model-based techniques are built upon predefined statistical models that capture the intrinsic characteristics of the \ac{ecg} waveform.
In contrast, non-parametric methods, such as the wavelet transform and \acs{emd}, steer clear of rigid model assumptions. They decompose the \ac{ecg} signal into different frequency components or intrinsic mode functions, facilitating the separation of noise. On the other end of the spectrum, deep neural network architectures harness vast amounts of data to learn an empirically optimal process for denoising an \ac{ecg} signal~\cite{chatterjee2020review, mir2021ecg, 10.1117/12.2660732, tripathi2021review}.

Deep learning-based approaches, specifically training \aclp{dnn} \acl{e2e} to minimize a loss function with vast datasets, have become powerful tools for various tasks~\cite{lecun2015deep}, including \ac{ecg} denoising~\cite{8902833}. Examples include using a recurrent neural network, as in~\cite{antczak2018deep}, and employing a fully convolutional denoising autoencoder, as in~\cite{8693790}. In~\cite{Fotiadou2020MultiChannelFE}, a multi-channel fetal \ac{ecg} denoising was considered based on deep convolutional neural networks. In~\cite{SinghP21}, a generative adversarial network architecture was considered, while in~\cite{RASTIMEYMANDI2022103275}, a \acl{dl} framework based on stacked cardiac cycle tensors was introduced.

A primary advantage of deep learning is its data-driven nature. When trained on diverse and comprehensive datasets, it can yield superior accuracy and robustness. The efficacy of these models is heavily influenced by the quality and diversity of the training data, highlighting the importance of meticulously curated datasets. However, these models also pose challenges. A significant constraint is their reliance on aggregated patient datasets. When trained with an \ac{mse} criterion, deep networks often exhibit a strong bias towards the mean. Consequently, in situations with noise where multiple plausible waveforms might have produced the observed data, these \ac{mse}-trained networks often output the posterior mean. This bias complicates the task of tailoring the model to an individual patient's unique \ac{ecg} characteristics. Furthermore, the complexity of these architectures requires substantial computational resources. Additionally, their "black box" nature introduces challenges in model interpretability.

Among the arsenal of effective \ac{ecg} denoising techniques, non-parametric methods have attracted considerable attention~\cite{kabir2012denoising}. \ac{emd} methods~\cite{10.2307/53161}, such as~\cite{weng2006ecg, lu2009model, zhang2020efficient}, stand out for their ability to decompose a signal into intrinsic mode functions (IMFs), with each IMF representing a specific oscillatory mode within the original \ac{ecg}. While EMD's decomposition provides a clear representation of the \ac{ecg}'s frequency components and facilitates effective noise separation, it can sometimes be sensitive to fluctuations, leading to mode mixing or the emergence of spurious modes. Additionally, wavelet transforms offer a robust multi-resolution analysis, enabling a precise time-frequency representation of signals~\cite{daubechies1992ten, shensa1992discrete, 6522142}. Methods, such as~\cite{sayadi2006ecg, alfaouri2008ecg, 5332617, georgieva2013denoising, singh2017denoising}, distinguish essential \ac{ecg} components from high-frequency noise and facilitate selective noise removal, preserving the integrity of the original \ac{ecg} waveform. While their computational efficiency makes them suitable for real-time clinical applications, their success heavily relies on the correct choice of wavelet and its parameters, and incorrect tuning can lead to \ac{ecg} distortion. Moreover, in the face of heavy noise, wavelets might not be fully effective.

Model-based techniques have emerged as a notable alternative for \ac{ecg} denoising. Central to this approach is the utilization of \ac{ss} models, complemented by various variations of the \ac{kf}~\cite{kalman1960new}, as in~\cite{sameni2007nonlinear, SayadiS08a, VullingsVB11, hesar2020adaptive, ouali2013ecg}. Local approximations of the \ac{ecg} waveforms have been explored through windowed \ac{ss} models in~\cite{WildhaberZJL18, Wildhaber19, WildhaberRWL20}, and via autoregressive models~\cite{evaristo2018mathematical, huang2019ecg}. Nevertheless, these approximations can sometimes fall short of capturing the intricate intra-heartbeat dynamics and the inherent quasi-periodicity between consecutive heartbeats, posing challenges for optimal denoising. 

A fundamental principle in numerous \ac{ecg} denoising methodologies is to model the signal's complex evolution by capitalizing on its quasi-periodicity. This principle is subsequently utilized as a prior belief in Bayesian filtering techniques~\cite{sarkka2013bayesian}, such as the \ac{kf}~\cite{sameni2007nonlinear}. For instance, \cite{VullingsVB11} omits intra-heartbeat variations and chooses instead to represent the evolution of consecutive heartbeats with an identity function. This approach is grounded in the premise that, in the absence of arrhythmia, two successive heartbeats, when centered around the R-peak, closely resemble each other. In essence, this method is equivalent to a weighted average of multiple heartbeats. While this strategy enhances the \ac{snr}, it also runs the risk of obscuring vital physiological dynamics. The research presented in~\cite{hesar2020adaptive} incorporates the \ac{em} algorithm to determine the evolution function. When combined with a bank of \ac{kf}s, this method efficiently targets both high and low-frequency noise. However, its application is limited by its reliance on a linear prior function, its constraint to filter signals of a fixed length, and its omission of the \ac{ecg}'s periodic information. In contrast,~\cite{McSharryCTS03} introduces non-linear priors using partial differential equation models~\cite{sameni2007nonlinear}. While innovative, these models frequently face challenges in capturing patient-specific variations. To address this issue,~\cite{ouali2013ecg} attempts to automatically fit model parameters using a \ac{ls} optimizer, based on several pre-recorded \aclp{hb}.
%
%

In this work, we propose a \ac{hkf} for \ac{ecg} denoising, which enhances signal quality without obscuring dynamic changes potentially linked to pertinent pathophysiology. Our \ac{hkf} is designed based on our innovative hierarchical \ac{ss} model, which describes the \ac{ecg} signal dynamics both within individual heartbeats and across consecutive heartbeats. Specifically, \ac{hkf} consists of an online learned structured evolution prior for a single heartbeat; a \ac{rts} intra-heartbeat smoother~\cite{rauch1965maximum} that harnesses this prior; and an inter-heartbeat \ac{kf}~\cite{kalman1960new} for denoising spanning multiple heartbeats. The online warm-up phase is meticulously designed to tackle challenges such as the highly patient-specific heartbeat shape and substantial noise variation resulting from myriad factors, ranging from equipment intricacies to room temperature variations. While one might conceptualize a typical \ac{ecg} signal shape, the reality is that there's considerable inter-variability among patients. Not only can the signal shape vary significantly between patients, but there's also intra-patient variability; the placement and orientation of electrodes can introduce alterations in the observed waveform. This renders the task of crafting a universal prior quite challenging. Crucially, \ac{hkf} doesn't require supervised pre-training and is inherently patient-adaptive, due to its online covariance estimation and its learned structured evolution prior. Yet, it preserves the transparent and interpretable nature of the \ac{kf}.

Our experimental study shows that the proposed \ac{hkf} effectively denoises \ac{ecg} signals, even in challenging setups, while retaining the subtle, clinically valuable structures within the signals. These attributes make it especially suited for medical and healthcare applications where a high degree of confidence and reliability is essential.
%
%

The remainder of this paper is organized as follows: Section~\ref{sec:sysModel} formulates the task and introduces the hierarchical \ac{ss} model. Section~\ref{sec:HKF} delves into the details of the proposed \ac{hkf}. Section~\ref{sec:ParamEst} elaborates on parameter estimation. Finally, Section~\ref{sec:EmpEval} presents our empirical study, demonstrating that the \ac{hkf} surpasses both \acl{mb} and \acl{dd} benchmarks.
%
%
\begin{figure}
\begin{subfigure}[a]{0.495\columnwidth}
\centering
\includegraphics[width=1\columnwidth]{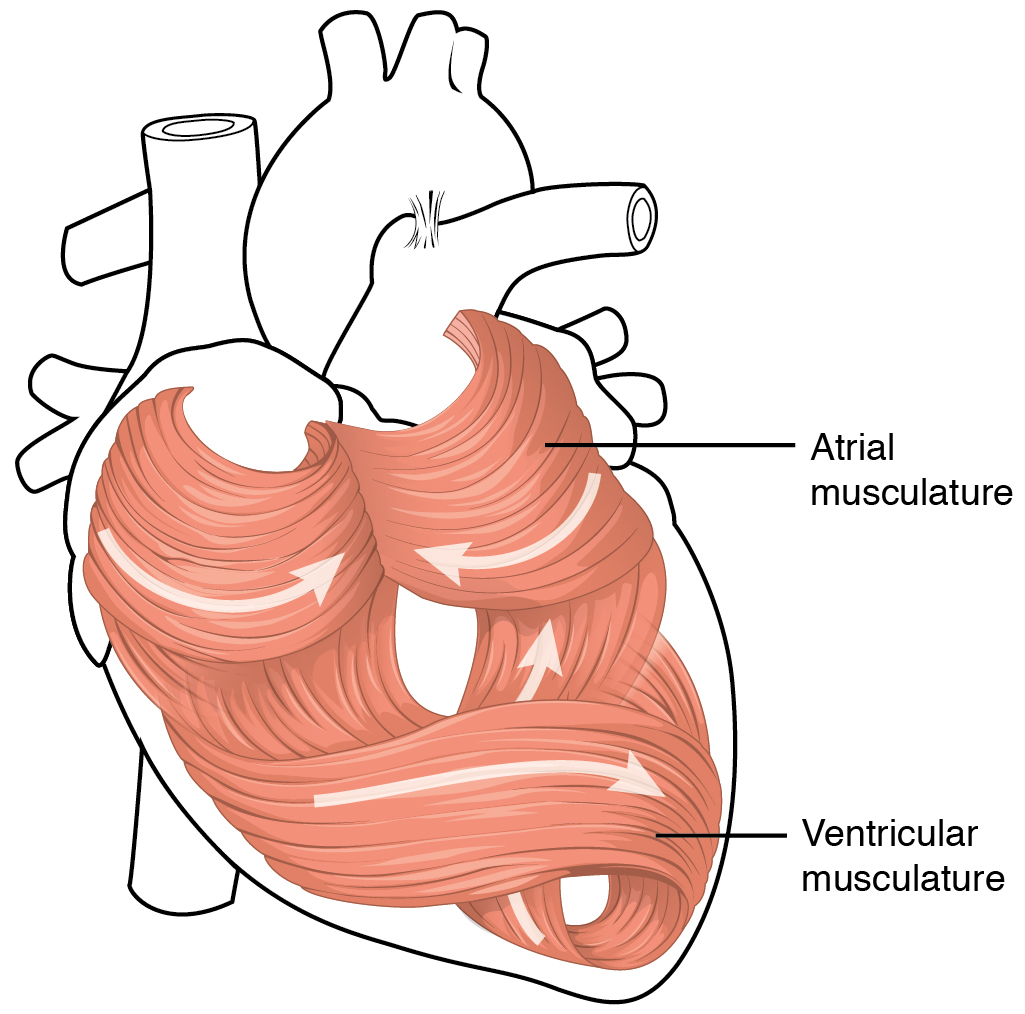}
\caption{Heart Musculature \cite{heartmuscle}} 
\label{fig:heart_muscle}
\end{subfigure}
\begin{subfigure}[a]{0.495\columnwidth}
\centering
\vspace{2.5cm}
\includegraphics[width=\columnwidth]
{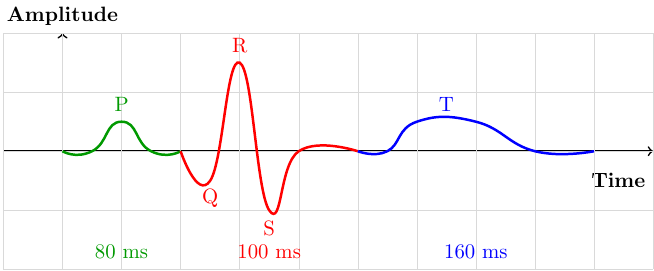}
\caption{An \ac{ecg} Waveform Illustration}   
\label{fig:ECG_WF}
\end{subfigure}
\caption{Illustration of a Heart and an \ac{ecg} Waveform}
\label{fig:Heart}
\end{figure}
\section{System Model and Task Formulation}\label{sec:sysModel}
In this section, we lay the groundwork for the subsequent derivation of the \ac{hkf}. We begin with a review of essential information on the \ac{ecg} signal. Following that, we delve into the task of multi-channel \ac{ecg} signal denoising. We conclude by presenting our unique hierarchical \ac{ss} model, which serves as the foundation for the \ac{hkf} design.
%
%
\subsection{The ECG Signal}
The heart is composed of two main types of muscle: the atrial muscle and the ventricular muscle, as illustrated in \figref{fig:heart_muscle}. During a \ac{hb}, these different muscles contract and relax in response to electrical impulses, which depolarize and repolarize the heart. These impulses propagate as an electrical field through the body and can be detected by electrodes on the skin. The voltage variation recorded over time is what we refer to as the \ac{ecg} signal. The typical \ac{ecg} signal comprises three segments: the \emph{P-wave}, the \emph{QRS complex}, and the \emph{T-wave}, as shown in \figref{fig:ECG_WF}. Each of these segments represents a different stage of heart contraction. The P-wave corresponds to the contraction of the atrial muscle; the QRS complex indicates ventricular depolarization; and the T-wave reflects ventricular repolarization. As its name suggests, the QRS complex consists of three smaller waves (Q-wave, R-wave, and S-wave) associated with the depolarization of the ventricular muscle. The entire QRS complex cycle takes about 100 $\msec$ \cite{310988}. While a typical \ac{ecg} signal depicts positive peaks for the P-wave, R-wave, and T-wave, the direction of these peaks can vary depending on the placement of the electrodes. The bandwidth of an \ac{ecg} signal usually falls within the 0.05-100 \si{\hertz} range \cite{allen1996assessing}.
%
%
\subsection{Multi-Channel ECG Denoising Task Formulation}
The electrical activity of the heart is observed and monitored by placing multiple electrodes on the human body and recording the noisy amplitudes as a vector time series.
Here, each electrode is referred to as a channel. The vector $\oihb{t}\in\greal^m$ denotes the observed noisy amplitudes across $m$ channels in a discrete-time index $t\in\gint$. The specific features of this recording, such as \ac{ecg} shape, amplitude, noise, and other artifacts, depend on the electrode and its placement on the body. We assume that $\oihb{t}$, the noisy recordings, originated from $\ihb{t}\in\greal^m$, multi-channel noiseless signals, that were then corrupted by \ac{agn} $\gvec{v}_t$. The \acs{ecg} denoising task is hereby defined as the reconstruction of $\ihb{t}$, the $m$ clean channels, from 
$\set{\oihb{i}}_{i=1}^t$, their corresponding past and current noisy observations, namely
\begin{equation}
\Psi:\set{\oihb{i}}_{i=1}^t\mapsto\hihb{t},
\hspace{0.275cm}
\Psi^\ast=\arg\min
\mathbb{E}{\norm{\hihb{t}-\ihb{t}}^2}.
\end{equation}
The denoiser, formulated as a mapping $\Psi$, is designed to minimize a cost function. A natural choice for this function is the \ac{mse}. 

The underlying ground truth heart activity is commonly described as a cardiac vector, a stochastic process in $\greal^3$, that captures the direction and magnitude of the electrical impulses as they propagate through the heart. Therefore we assume that $\ihb{t}$, describing the $m$ channels, is intra-correlated, and originates from the same \emph{hidden} \acl{gt} activity signal
This assumption will be used in our hierarchical system model as shown next.  
%
%
%
\begin{figure}
\centering
\includegraphics[width=0.75\columnwidth]{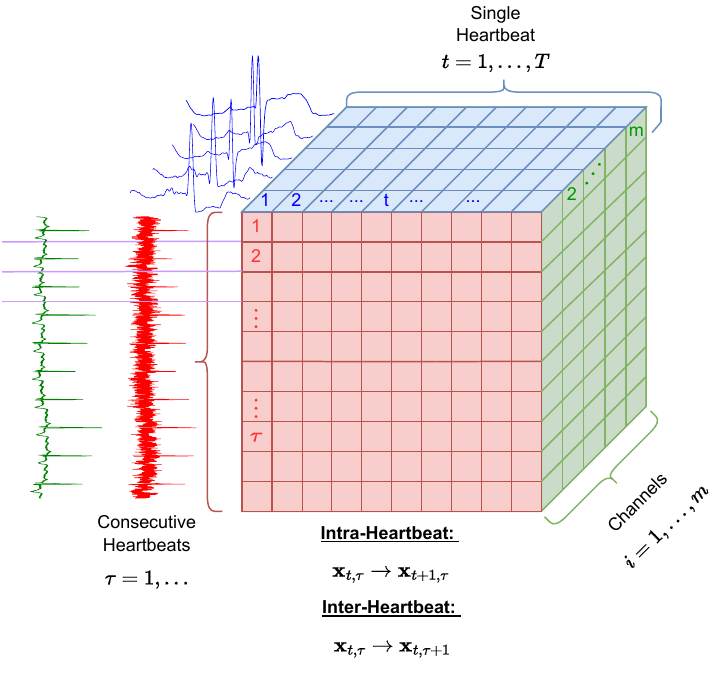}
\caption{Hierarchical System Model as a Tensor}
\label{fig:hierachical_sysmodel}
\end{figure}
\subsection{Hierarchical System Model}
The relationship between the observed  \ac{ecg} signal $\oihb{t}$ to its corresponding noiseless instance $\ihb{t}$, can be modeled as a dynamical system. A canonical way to model dynamical systems in \acl{dt} is by using \ac{ss} models \cite{bar2004estimation}. \ac{ss} models blend well with numerous filtering and smoothing techniques, and can accommodate the integration of both physical and \acl{dd} learned models. To exploit the periodicity, which plays a key role in the \ac{ecg} signal, we model the signal dynamics using a hierarchical \ac{ss} model. 

In the following, we make a reasonable assumption that individual \acp{hb} can be accurately segmented. Therefore, we divide the signal into periodic segments of length $\thor$, where each such segment represents a single \ac{hb}. Our model draws inspiration from the decimated components decomposition method, which is used to represent periodic systems and cyclostationary signals \cite{giannakis1998cyclostationary} as multivariate (tensor) stationary ones. The dynamics within a single \ac{hb} are described using an \emph{intra}-\ac{hb} (internal) \ac{ss} model, while the dynamics between two consecutive \aclp{hb} are modeled using an \emph{inter}-\ac{hb} (external) \ac{ss} model, as detailed next. 
%
%
\subsection{Intra Heartbeat Modeling}
The \emph{intra}-\ac{hb} \ac{ss} model defines the  
{dynamics} within a {single} \ac{hb} with index $\tau$. For each time step $t$ within the period $\tthor\triangleq\set{1,..,\thor}$ of length $\thor$, the model is given by:
\begin{subequations}\label{eq:ss_intra}
\begin{align}
\ihb{\tau, t} &= \ievol{\ihb{\tau, t-1}}+\gvec{e}_{\tau,t},
&\gvec{e}_{\tau,t}&\sim\gnormal{0,\ievolcov{ t}},\\
\label{eq:intra_obs}
\oihb{\tau,t} &= \ihb{\tau,t}+\gvec{v}_{\tau,t},
&\gvec{v}_{\tau,t}&\sim\gnormal{0,\iobscov{t}}.
\end{align}
\end{subequations}
The matrices $\ievolcov{t}$ and $\iobscov{t}$ represent the time-varying intra-evolution (process) and observation covariance of a Gaussian distribution, respectively. They capture the correlation across multiple channels. Specifically, $\ievolcov{t}$ accounts for the correlation arising from the fact that these channels originate from the same foundational heart activity. In contrast, $\iobscov{t}$ captures the correlation attributed to shared measuring effects: patient-related factors such as breathing and movement, measuring-device-related factors such as quality and age, and environmental factors, including electromagnetic interference and ambient room temperature. As a result, these matrices are assumed to be unknown and are not restricted to be diagonal.

We denote the $\tau$-th \ac{hb} and its noisy observation as the matrices $\hb{\tau}$ and $\ohb{\tau}$, respectively, namely,
\begin{subequations}
\begin{align}
\hb{\tau}&=
\sbrackets{\ihb{\tau,1},\hdots,\ihb{\tau,\thor}}
\in\greal^{m\times\thor},\\
{\ohb{\tau}}&=
\sbrackets{\oihb{\tau,1},\hdots,\oihb{\tau,\thor}}
\in\greal^{m\times\thor}.
\end{align}
\end{subequations}
%
%
\subsection{Inter Heartbeat Modeling}
The \emph{inter}-\ac{hb} \ac{ss} model defines the 
evolution between two {consecutive} \acp{hb}, labeled with indices $\tau-1$ and $\tau$, namely:
\begin{subequations}\label{eq:ss_inter}
\begin{align}
\ihb{\tau, t} &= \eevol{\ihb{\tau-1, t}}+\eevoln{\tau,t},
&\eevoln{\tau,t}&\sim\gnormal{0,\eevolcov{\tau, t}},\\
\tilde{\gvec{y}}_{\tau,t} &= \ihb{\tau,t}+{\boldsymbol{\nu}}_{\tau,t},
&{\boldsymbol{\nu}}_{\tau,t}&\sim\gnormal{0,\eobscov{\tau,t}}.
\end{align}
\end{subequations}
Based on the periodicity of the cardiac vector, here, we assume that two consecutive \acp{hb} closely resemble each other, therefore, the inter-state evolution is defined as the identity mapping, i.e., $\eevol{\ihb{}}=\ihb{}$. The matrices $\eevolcov{\tau, t}$ and $\eobscov{\tau, t}$ represent the inter-evolution (process) and observation covariance respectively. These matrices capture the multi-channel correlation, and they adhere to the same assumptions outlined for the inter-HB model. 
\begin{equation}
\gvec{Y}_{\tau}=
\sbrackets{\tilde{\gvec{y}}_{\tau,1},\hdots,\tilde{\gvec{y}}_{\tau,\thor}}
\in\greal^{m\times\thor}.    
\end{equation}
An illustrative overview of the system model can be found in \figref{fig:hierachical_sysmodel}.
%
%
\section{HKF - Hierarchical Kalman Filter }\label{sec:HKF}
Our custom-designed \ac{hkf} exploits the proposed hierarchical \ac{ss} model to perform patient-dependent denoising.
%
%
\subsection{Overview and Design Rationale}
\label{ssec:overview}
There are two main phases to the algorithm:
\begin{enumerate}[label=\color{blue}P\arabic*]
\item A short \emph{online warm-up} phase.\label{p1}
\item A \emph{signal-denoising processing} phase.\label{p2}
\end{enumerate}
In~\ref{p1}, the \emph{online warm-up phase}, we learn the internal parameters of the intra-\ac{hb} \ac{ss} model~\eqref{eq:ss_intra}, specifically: the prior signal evolution model and the noise covariance matrices. The specific \ac{hb} shape varies considerably among patients and can be influenced by numerous factors, ranging from physiological to equipment-related and ambient factors. Consequently, pre-training an algorithm on a \acl{ds} from multiple patients results in learning an average heartbeat waveform. To avoid this, our algorithm is designed to adapt to individual patients, learning the signal model in an \emph{unsupervised} manner. In particular, we employ Taylor \ac{ls}, detailed in~\ssecref{sec:online_prior}, for the evolution function approximation. Meanwhile, the unknown noise covariance matrices are estimated using a variant of the \ac{em} algorithm, as elaborated upon in~\ssecref{sec:online_covariance}.

Next, in phase \ref{p2}, the \emph{signal-denoising processing} phase, each new \ac{hb} $\hb{\tau}$ is denoised by fusing $\ohb{\tau}$, its noisy observation, with $\set{\ohb{\tau'}}_{\tau'=1}^{\tau-1}$, which encapsulates all previous information. Namely:
\begin{equation}
\Psi: \set{\ohb{\tau'}}_{\tau'=1}^{\tau} \mapsto
\hhb{\tau}
\end{equation}
The fusion process is conducted efficiently through a recursive update, which fuses the new observation with the previous posterior, $\hhb{\tau}$, acting as a sufficient statistic for the accumulated information. This approach is underpinned by the SS models outlined earlier. Namely:
\begin{equation}
\Psi: \hhb{\tau-1}, \ohb{\tau}
\mapsto
\hhb{\tau}
\end{equation}
Here, for enhanced clarity, we use $\hhb{\tau}$ to denote $\hhb{\tau\given{\tau}}$, which represents the mean of the posterior distribution, given $\tau$ \acp{hb}.

In practice, the denoising is implemented in two steps as in the following:
\begin{enumerate}[label=\color{blue}D\arabic*]
\item Intra-\ac{hb} Kalman smoothing. \label{d1}
\item Inter-\ac{hb} Kalman filtering. \label{d2}
\end{enumerate}
In \ref{d1}, the \ac{rts} smoother~\cite{rauch1965maximum, sarkka2013bayesian} is used to compute $\thb{\tau}$, an intermediate denoised version, from $\ohb{\tau}$, in a stand-alone manner, based on 
the intra-\ac{hb} \ac{ss} model \eqref{eq:ss_intra}, given the learned prior parameters. More specifically:
\begin{equation}
\Tilde{\Psi}:
\ohb{\tau}\mapsto
\thb{\tau}=
\sbrackets{\hihb{\tau,1\given{\thor}},\hdots,\hihb{\tau,\thor\given{\thor}}}
\in\greal^{m\times\thor}.
\end{equation}
Here, the notation $\hihb{\tau,t\given{\thor}}$ denotes the smoothed version of sample $t$ in \ac{hb} $\tau$ given all $\thor$ samples from the entire $\ac{hb}$. This stage is detailed in \ssecref{sec:first_stage_smoothing}.

In \ref{d2}, the \ac{kf} is used to compute $\hhb{\tau}$, the fully denoised posterior \ac{hb}, by fusing $\gvec{Y}_{\tau}=\thb{\tau}$, i.e., the stand-alone smoothed version, with $\hhb{\tau-1}$, the fully denoised posterior of the previous heartbeat. This stage is based on the inter-\ac{hb} \ac{ss} model \eqref{eq:ss_inter}, with an adaptive covariance estimation. More specifically: 
\begin{equation}
\hat{\Psi}:
\hhb{\tau-1}, \thb{\tau}
\!\mapsto\!
\hhb{\tau}\!=\!
\sbrackets{\hihb{\tau\given{\tau},1\given{\thor}},\ldots,
\hihb{\tau\given{\tau},\thor\given{\thor}}}.
\end{equation}
This operation represents a batch of $\thor$ parallel \acp{kf}. In this context, the additional notation $\tau\given{\tau}$ denotes the update of the smoothed version, incorporating information from the current and all previous \acp{hb}. Further details are provided in ~\ssecref{sec:second_stage_filtering}.

\begin{figure}
\centering
\includegraphics[width=0.7\columnwidth]{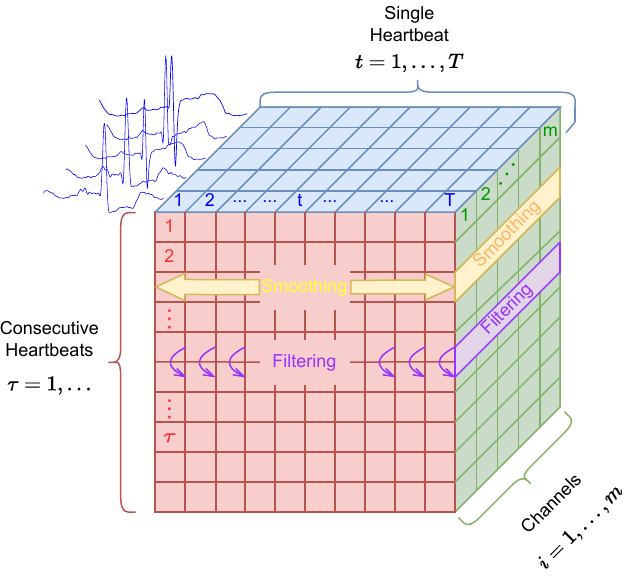}
\caption{Processing Phase - Smoothing and Filtering}
\label{fig:smoothing_filter_cube}
\end{figure}

\begin{figure}
\centering
\includegraphics[width=1\columnwidth]{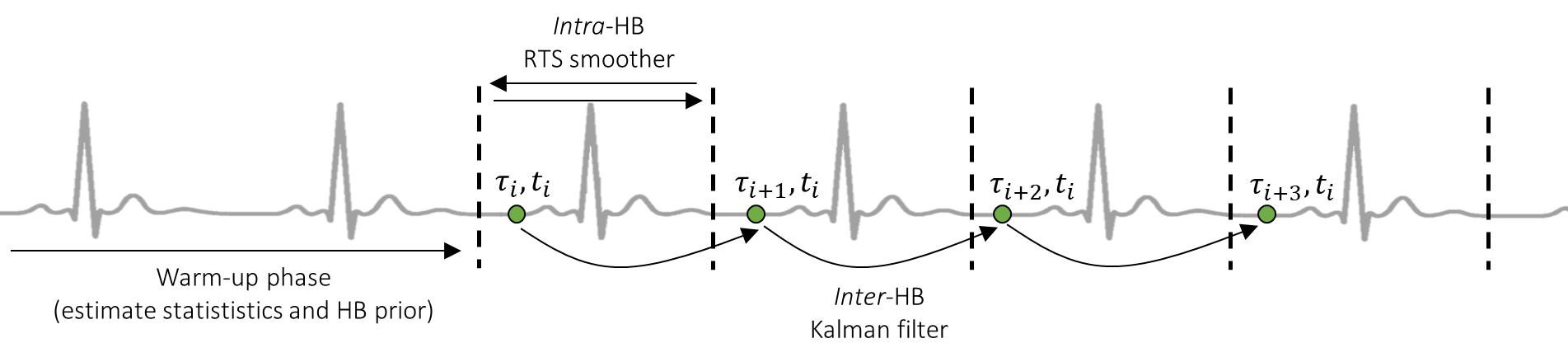}
\caption{HKF Overview}
\label{fig:HKFoverview}
\end{figure}

\figref{fig:smoothing_filter_cube} visualizes the high-level structure of the processing phase, while \figref{fig:HKFoverview} is a high level description of our \ac{hkf} algorithm.
%
%

%
%

%
%
%
\subsection{First Stage Smoothing}\label{sec:first_stage_smoothing}
In the processing phase, the first step, \ref{d1}, involves implementing a \ac{ks} using the \ac{rts} algorithm~\cite{rauch1965maximum}. This \ac{ks} operates based on the intra-\ac{hb} \ac{ss} model parameters \eqref{eq:ss_intra}: 
$\hat{\gvec{f}}_t$, $\hat{\ievolcov{}}_t$, and $\hat{\iobscov{}}_t$, learned during the warm-up phase. Given $\ohb{\tau}$, the newly observed \ac{hb} with index $\tau$, we employ the \ac{ks}, denoted as $\Tilde{\Psi}$, to obtain an instantaneous estimate of the denoised \ac{hb}, denoted as $\hhb{\tau}$.

The \ac{rts} smoother encompasses two-steps:
\begin{enumerate}[label=\color{blue}KS\arabic*]
\item A \emph{forward} pass step, i.e., a \ac{kf}.\label{ks1}
\item A \emph{backward} pass step. \label{ks2}
\end{enumerate}
The forward pass, \ref{ks1}, is defined by two phases: \emph{prediction} and \emph{update}\footnote{For brevity we ignore the inter-\ac{hb} index $\tau$.} The \emph{prediction} phase, is given by state prediction:
%
%
\begin{equation}\label{eqn:predict1}
\hihb{t\given{t-1}} = 
\ievol{\hihb{t-1\given{t-1}}},
\quad
\ievolErr{t\given{t-1}}=
\ievolErr{t-1\given{t-1}}+\ievolcov{t},
\end{equation}
where $\ievolErr{t\given{t-1}}$, the prior covariance, is computed using an identity matrix, as defined in~\eqref{eq:jacob}, and by innovation prediction:
\begin{equation}
\hoihb{t\given{t-1}} =
\hihb{t\given{t-1}},
\quad
\Delta\oihb{t}=\oihb{t}\!-\!\hoihb{t\given{t-1}},
\quad
\iobsErr{t}=
\ievolErr{t\given{t-1}}+\iobscov{t}.   
\end{equation}
The \emph{update} phase is given by:
%
%
\begin{equation}\label{eqn:update}
\hihb{t\given{t}}= 
\hihb{t\given{t-1}}\!+\!\iKgain_{t}\cdot\Delta\oihb{t},
\quad
\ievolErr{t\given{t-1}}-\iKgain_{t}\cdot\iobsErr{t}\cdot\iKgain_{t}^\top.
\end{equation}
Here, $\iKgain_{t}$ is the \acl{kg}, given by:
\begin{equation}
\iKgain_{t}=\ievolErr{t\given{t-1}}\cdot\iobsErr{t}^{-1}
\end{equation}
The {backward} pass, \ref{ks2}, of the \ac{rts} smoother is given by:
\begin{equation}
\hihb{t\given{\thor}}=
\hihb{t\given{t}} + \iSgain_{t}\cdot\Delta\ihb{t}, 
\quad
\ievolErr{t\given{\thor}}=\ievolErr{t|t} - \iSgain_t\cdot \Delta\ievolErr{t+1} \cdot \iSgain_t.
\end{equation}
\begin{equation}
\Delta\ihb{t}=\hihb{t+1\given{\thor}}-\hihb{t+1\given{t}},
\quad
\Delta\ievolErr{t+1}=\ievolErr{t+1\given{\thor}}-\ievolErr{t+1\given{t}}
\end{equation}
Here, $\iSgain_{t}$, the smoothing gain, is given by: 
\begin{equation}
\iSgain_t=\ievolErr{t\given{t}}
\ievolErr{t+1\given{t}}^{-1}.    
\end{equation}
The output of the smoothing step is 
\begin{equation}
\thb{\tau}= \Tilde{\Psi}\brackets{\ohb{\tau}} =
\sbrackets{\hihb{\tau,1\given{\thor}},\hdots,\hihb{\tau,\thor\given{\thor}}}.
\label{eqn:smoothout}
\end{equation}
We assume that the estimated distribution for the \emph{instantaneously} denoised \ac{hb} is given by 
\begin{equation}
\gvec{x}_{\tau, t\given{\thor}}\sim
\gnormal{\hihb{\tau, t\given{\thor}}, \gmat{P}_{\tau, t\given{\thor}}}
\end{equation}
This distribution is significant for the subsequent filtering stage detailed in~\ssecref{sec:second_stage_filtering}. We will use $\hat{\gvec{x}}_{t\given{\thor}}$, the posterior mean, and $\gmat{P}_{\tau, t\given{\thor}}$, the posterior error covariance estimate, from the initial stage as observations and observation noise for the second stage, respectively. This approach not only eliminates the need for an additional estimation step but also offers the optimal estimate of the observation noise matrices, under the assumption that all previously estimated values are optimal.
%
%
\subsection{Second Stage Filtering} \label{sec:second_stage_filtering}
In the second step, \ref{d2}, of the processing phase, we employ a \ac{kf} based on the inter-\ac{hb} \ac{ss} model~\eqref{eq:ss_inter}. Given the immediate estimate of the current \ac{hb}, denoted as $\thb{\tau}$, and the posterior estimate of the preceding \ac{hb} accounting for its entire past $\hhb{\tau-1}$, we produce a posterior estimate for the current \ac{hb} $\hhb{\tau}$. This filter effectively unfolds to $\thor$ \acp{kf} operating simultaneously along the $\tau$-axis:
\begin{equation}
\Psi^*_t: 
\hat{\gvec{x}}_{\tau, t\given{\thor}}, 
\hat{\gvec{x}}_{\tau-1\given{\tau-1}, t\given{\thor}}
\mapsto
\hat{\gvec{x}}_{\tau\given{\tau}, t\given{\thor}}.
\end{equation}
In the end, we leverage the temporal correlation between two successive \acp{hb}, $\tau-1$ and $\tau$, as outlined in~\eqref{eq:ss_inter}. For every $t\in\tthor$ and given the estimated covariance matrices, we implement $\gscal{T}$ independent (and parallel) \acp{kf}. This procedure is articulated in the ensuing steps; note that for simplicity, we occasionally omit the intra-\ac{hb} index $t$.
\begin{enumerate}[label=\color{blue}KF\arabic*]
\item Prior error and innovation covariance: \label{kf1}
\begin{equation}\label{eq:kf1}
\eevolErr{\tau\given{\tau-1}} =
\eevolErr{\tau-1}+
\heevolcov{\tau},
\quad
\eobsErr{\tau}=
\eevolErr{\tau\given{\tau-1}}+\heobscov{\tau}.
\end{equation}
\item The \emph{update} equations: \label{kf2}
\begin{equation}
\hspace{-0.25cm}
\hhb{\tau}=
\hhb{\tau-1}+\Kgain_{\tau}\cdot\Delta,
\quad
\eevolErr{\tau}=
\eevolErr{\tau\given{\tau-1}}-
\Kgain_{\tau}\cdot{\eobsErr{\tau}}\cdot\Kgain^{\top}_{\tau}.    
\end{equation}
\end{enumerate}
Here, $\Delta$ is the innovation, and $\Kgain_{\tau}$ is the Kalman gain:
\begin{equation}\label{eq:kf21}
\Delta=\thb{\tau}-\hhb{\tau-1}=
\brackets{\hat{\gvec{x}}_{\tau, t\given{\thor}}- 
\hat{\gvec{x}}_{\tau-1\given{\tau-1}, t\given{\thor}}}_{\forall t},
\end{equation}
\begin{equation}
\Kgain_{\tau}=
\eevolErr{\tau\given{\tau-1}}\cdot\eobsErr{\tau}^{-1} 
\end{equation}
To successfully deploy the \ac{kf} the set of inter-\ac{hb} covariance matrices, i.e.,  
\begin{equation}
\set{\eevolcov{\tau, t}, \eobscov{\tau, t}}_{t=1}^{\thor} 
\end{equation}
should be estimated on the fly before the filtering step. The estimation of the observation noise is provided in \ssecref{sec:inter_obs_est}, while the estimation of the process noise is provided in \ssecref{sec:inter_evol_est}.
%
%
%
%
\subsection{Signal Pre-Processing}\label{sec:pre_processing}
The accurate detection of the various peaks and intervals within an \ac{ecg} waveform, such as R-peaks and QT-interval, often provides vital clinical information. Over the years, a multitude of methods have been developed to address this complex task. Classical approaches typically utilize the time-derivatives of the recorded signal, applying thresholding on parameters such as slew-rate~\cite{christov2004real} or directly to the derivative~\cite{sathyapriya2014analysis}. Some methods attempt to fit local \ac{ss} models to a given signal, then detecting peaks by observing sudden changes in the model using a log-cost ratio~\cite{waldmann2022onset}.

\begin{figure}
\centering
\includegraphics[width=0.65\columnwidth]{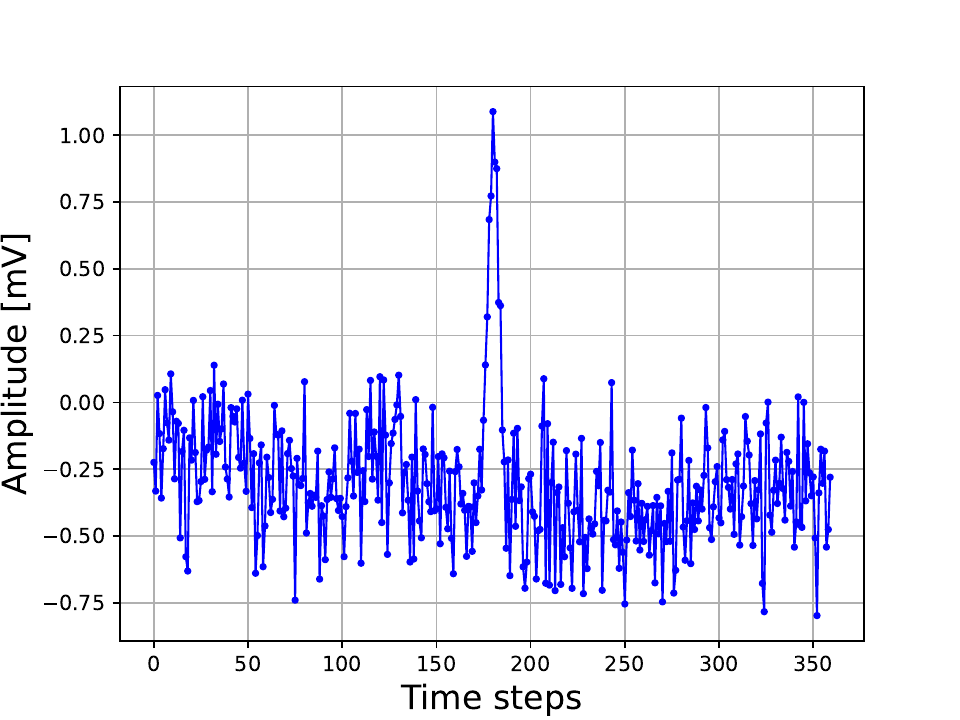}
\caption{Example of noisy observation $\ohb{\tau}$ after pre-processing.}
\label{fig:centered_obs}
\end{figure}

Our HKF algorithm operates under the assumption that each \ac{hb}, represented as $\hb{\tau}$, can be isolated during a pre-processing phase and subsequently processed independently. The central idea hinges on detecting R-peaks and processing a set of samples centered around them. The width of this processing window is proportional to the sampling frequency of the utilized ECG recording device (e.g. $f_s = 360 \si{\hertz} \rightarrow$ translates to 360 data points per \ac{hb}). A vast array of off-the-shelf peak-detection algorithms exist, ranging from wavelet transform-based methods like~\cite{fard2008novel, zou2014energy}, to \ac{nn} based techniques such as~\cite{laitala2020robust}. For our empirical studies, we leveraged the python bioSPPy library~\cite{biosppy}, which is based on~\cite{engelse1979single, lourencco2012real}. An example of a pre-processed observation is illustrated in~\figref{fig:centered_obs}.
%
%
\subsection{Signal Post-Processing}\label{sec:post_processing}
\begin{figure}
\hspace{-0.75cm}
\begin{subfigure}[a]{0.495\columnwidth}
\includegraphics[width=1.25\columnwidth]{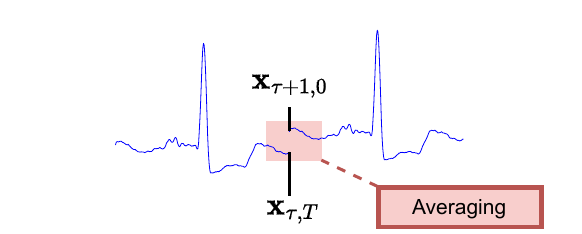}
\end{subfigure}
\hspace{-0.45cm}
\begin{subfigure}[a]{0.495\columnwidth}
\includegraphics[width=1.25\columnwidth]
{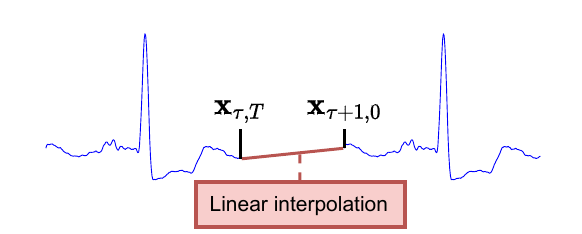}
\end{subfigure}
\caption{Reconstruction of Continuous \ac{ecg}}
\label{fig:stiching}
\end{figure}

After successfully filtering an individual HB, it can be seamlessly integrated with previously filtered HBs to form a continuous signal. Given the fixed window length \(T\) applied during pre-processing, there are two potential scenarios that can arise based on the \ac{hr}: Overlaps (denoted as \ref{pp1}) and Gaps (denoted as \ref{pp2}).
\begin{enumerate}[label=\color{blue}PP\arabic*]
\item Overlaps: \label{pp1}
In this scenario, a weighted average of the reconstructed version can be employed as the estimated continuous signal. 
%
%
\item Gaps: \label{pp2}
For gaps between heartbeats, linear interpolation is chosen to bridge the gap between the end of the \ac{hb} with index $\tau$ and the beginning of the subsequent \ac{hb} with index $\tau+1$. This approach is justified as there's typically no significant heart activity between these two periods, and the region usually consists predominantly of white noise.
%
%
\end{enumerate}
A visual representation of these two methodologies can be referenced in \figref{fig:stiching}.

%
%
\section{Online Prior Learning and Parameter Estimation}\label{sec:ParamEst}
In this section, we delve into the process of parameter learning. Specifically, we discuss how to determine the parameters for the intra-\ac{hb} Kalman smoothing (\ref{d1}) during the online \emph{warm-up} phase (\ref{p1}). We will also cover the learning of parameters for the inter-\ac{hb} Kalman filtering (\ref{d2}).
%
%
\subsection{Online Learned Taylor Priors}\label{sec:online_prior}
The primary objective of the online \emph{warm-up} phase (\ref{p1}) is to determine a patient-specific characterization of the intra-\ac{hb} \ac{ss} model \eqref{eq:ss_intra} parameters, with a particular focus on the \ac{ecg} waveform signal evolution model. Given the complexities of the \ac{ecg} signal, identifying a closed-form function that can accurately depict its evolution is challenging. As a solution, we adopt a \emph{Taylor} function approximation as our point-wise evolution model. The Taylor model provides insight into how a function behaves near a specific point, making it suitable for this purpose. Owing to the periodic nature of the \ac{ecg} signal, characterizing such a prior for one period, $\tau$, ensures its applicability across all periods.
\begin{equation}
\ihb{\tau, t}=\ievol{\ihb{\tau, t-1}}=\ihb{\tau, t-1}+\Delta\statevec{\tau, t}, 
\quad
t\in\tthor
\end{equation}
In this context, $\gvec{x}$ is a function of  $t$. Hence, for a sufficiently small time discretization interval, $\Delta t$, we can approximate $\Delta\statevec{\tau, t}$ using a finite Taylor expansion, provided that $K$ is sufficiently large.
\begin{equation}\label{eq:taylor_short}
\Delta\statevec{\tau, t}=
\statevec{\tau, t}-
\statevec{\tau, t-1}\simeq 
\sum_{k = 1}^{K} \frac{d^k}{dt^k}\statevec{\tau, t-1} \cdot\frac{\Delta t^k}{k!}=
\gvec{F}_t\cdot\phi\brackets{\Delta t}.
\end{equation}
Here, $\phi\in\mathbb{R}^{K\times 1}$ is a vector of a basis functions, namely:
\begin{equation}
\phi^\top\brackets{\Delta t}=  
\begin{pmatrix}
\frac{\Delta t^1}{1!}, \cdots,  \frac{\Delta t^K}{K!}
\end{pmatrix}, 
\end{equation}
and $\gvec{F}_t\in \mathbb{R}^{m \times K}$ is designated as a matrix of coefficients, formulated as:
\begin{equation}
 \gvec{F}_t = \begin{pmatrix}
\frac{d^1}{dt^1}\statevec{t-1,\tau}, \cdots, 
\frac{d^K}{dt^K}\statevec{t-1,\tau}
\end{pmatrix}.
\end{equation}
Considering this, the Jacobian matrix, $\mathcal{J}_{\gvec{f}}$, which is necessary for integrating the evolution into an \ac{ekf}, simplifies to an identity matrix. Specifically:

\begin{equation}\label{eq:jacob}
\mathcal{J}_{\gvec{f}}=
\frac{\gscal{d}\statevec{\tau, t}}{\gscal{d}\statevec{\tau, t-1}}=
\frac{\gscal{d}\ievol{\ihb{\tau, t-1}}}{\gscal{d}\statevec{\tau, t-1}}=
\gvec{I}_{m \times m}
\end{equation}
This comes in handy in equation \eqref{eqn:predict1}.
%
%
%
\subsection{Learning a Taylor Approximation}
The matrix $\gvec{F}_t$, which remains to be determined, can be learned in an unsupervised manner from a set of noisy observations of size $N$. This set is given by: 
\begin{equation}
\set{\Delta\stateobs{\tau_i, t}=\stateobs{\tau_i, t}-
\stateobs{\tau_i, t-1}}_{i=1}^N, 
\end{equation}
The objective is to minimize a specific \ac{ls} loss function, specified as:
\begin{equation}
\mathcal{L}\brackets{\gvec{F}_t}=
\frac{1}{N}\cdot
\sum_{i=1}^N\norm{\Delta\stateobs{\tau_i, t}-\gvec{F}_t\cdot\phi}^2,
\end{equation}
where $\hat{\gvec{F}}_t$ represents the minimized value:
\begin{equation}
\hat{\gvec{F}}_t=\arg\min\set{\mathcal{L}\brackets{\gvec{F}_t}}
\quad
\gvec{F}_t\in \mathbb{R}^{m \times K}.   
\end{equation}
This optimization problem offers a clear, closed-form solution. It can be efficiently computed through a single multiplication of an online-generated column vector $\Delta\overline{\gvec{y}}_t$ by a constant row vector $\gvec{g}$, namely:
\begin{equation}
\hat{\gvec{F}}_t=
\Delta\overline{\gvec{y}}_t
\cdot\gvec{g},
\end{equation}
where 
\begin{equation}
\Delta\overline{\gvec{y}}_t=
\frac{1}{N}\cdot\sum_{i=1}^N 
\Delta\stateobs{\tau_i, t}, 
\quad
\gvec{g}=\phi^\top\cdot\brackets{\phi\cdot\phi^\top}^{-1}.
\end{equation}
The effectiveness and reliability of our estimator are influenced by the number of observations used for its computation. This, in turn, relies on the number of \acp{hb} utilized to gather data. An additional objective is to ensure the smooth evolution of our \ac{ecg} waveform. To achieve this smoothness, we can aggregate multiple observations in the close proximity of \acl{ts} of \acl{ts} and adapt our loss function accordingly:
\begin{equation}\label{eq:opt_weighted_L2}
\mathcal{L}_{\alpha}\brackets{\gvec{F}_t}=
\frac{1}{N}\cdot
\sum_{i=1}^N
\sum_{j=-M}^M
\alpha_j\cdot
\norm{\Delta\stateobs{\tau_i, t+j}-\gvec{F}_t\cdot\phi}^2. 
\end{equation}
Specifically, we employ a window spanning $2M+1$ with decaying weights $\alpha_j$ that sums up to one:
\begin{equation}
\sum_{j=-M}^M
\alpha_j=1.
\end{equation}
The weighted \ac{ls} optimization problem in \eqref{eq:opt_weighted_L2} also admits a straightforward solution, given by:
\begin{equation}
\hat{\gvec{F}}_{\alpha,t}=
\Delta\overline{\overline{\gvec{y}}}_t
\cdot\gvec{g}, 
\quad
\Delta\overline{\overline{\gvec{y}}}_t=
\frac{1}{N}\cdot
\sum_{i=1}^N
\sum_{j=-M}^M
\alpha_j\cdot
\Delta\stateobs{\tau_i, t+j}.
\end{equation}
%
%
\subsection{Online Covariance Learning}\label{sec:online_covariance}
In the second stage of our proposed \emph{warm-up} phase \ref{p1}, the primary objective is to estimate the missing covariance matrices for the intra-\ac{hb} \ac{ss} model \eqref{eq:ss_intra}. Specifically, we aim to determine the intra-evolution covariance, $\ievolcov{t}$, and the intra-observation covariance, $\iobscov{t}$. To achieve this, we employ the \ac{em} algorithm. The \ac{em} algorithm, an iterative\footnote{For brevity we ignore the iteration index} method, has been frequently adapted for the task of parameter estimation in \ac{ss} models~\cite{ghahramani1996parameter, dauwels2009expectation, SophoclesJ.Orfanidis2018}.

The \ac{em} algorithm operates by alternating between the E-step and the M-step. During the E-step, we compute the conditional expectation. For our application, this involves calculating the posterior moments using the \ac{rts} smoother for each \acl{ts} $t\in\tthor$. Specifically, from the \ac{rts} smoother, we derive the first-order moment $\hihb{t\given{\thor}}$, the covariance matrix $\ievolErr{t\given{\thor}}$, and the backward smoothing gain $\iSgain_{t}^\top$. With these values, we can then determine the posterior second-order moments, namely:
\begin{equation}
\gvec{X}^{\gscal{II}}_{t}=
\hihb{t\given{\thor}}\cdot\hihb{t\given{\thor}}^\top
+\ievolErr{t\given{\thor}},
\quad
\gvec{Y}^{\gscal{II}}_{t}=\oihb{t}\cdot\oihb{t}^\top,    
\end{equation}
and the posterior correlations:
\begin{subequations}
 \begin{align} 
\gvec{XY}_{t,t}&=\hihb{t\given{\thor}}\cdot\oihb{t}^\top,\\
\gvec{XX}_{t,t-1}&=\hihb{t\given{\thor}}\cdot\hihb{t-1\given{\thor}}^\top + \ievolErr{t\given{\thor}}\cdot \iSgain_{t-1}^\top.
\end{align}     
\end{subequations}
%
%
In the M-step, we employ \acl{mle}, drawing upon the results of the previous step, to derive an instantaneous estimate for the unknown covariance matrices, as outlined next:
\begin{subequations}
\begin{align}
\tievolcov{t} &= 
\gvec{X}^{\gscal{II}}_{t}-2\cdot\gvec{XX}_{t,t-1}+\gvec{X}^{\gscal{II}}_{t-1},\\
\tiobscov{t}&= 
\gvec{Y}^{\gscal{II}}_{t}-2\cdot\gvec{YX}_{t,t}+\gvec{X}^{\gscal{II}}_{t}.
\end{align}   
\end{subequations}
Given the distinct shape of the \ac{hb}, we operate under the assumption that the observation noise statistic remains invariant within a single \ac{hb}; that is, it stays consistent over the timescale of a \ac{hb}. Consequently, by averaging the instantaneous estimators across the entire \ac{hb}, we obtain a single estimate that holds valid for all \aclp{ts}, as illustrated next:
\begin{equation}
\hat{\iobscov{}}_t = 
\hat{\iobscov{}} =  
\frac{1}{\thor}\cdot\sum_{t = 1}^{\thor}\Tilde{\iobscov{}}_t.
\end{equation}
While our previous assumption applies to the observation noise statistic, it does not hold true for the process noise statistic. We assume that this statistic is not invariant within a single \ac{hb} and exhibits temporal correlations over brief periods. Consequently, we take an average of the instantaneous estimators within a short time window of length $\tilde{L}=L_1+L_2+1$, obtaining:
\begin{equation}\label{eq:smooting_window}
\hat{\ievolcov{}}_t = 
\frac{1}{L_1 + L_2 + 1}\cdot\sum_{\ell= -L_1}^{L_2} \Tilde{\ievolcov{}}_{t + \ell} .
\end{equation}
To get a more robust and accurate estimate, we can exploit inter-\ac{hb} correlations and unfold the \ac{em} algorithm over multiple \acp{hb}. Given a sequence of \acp{hb}, the most straightforward way is to apply the \ac{em} algorithm on each \ac{hb} independently, and then average the single \ac{hb} based estimators to a multi-\ac{hb} based estimator. The second alternative, which is more aligned with online learning, is to run a single \ac{em} iteration on the data from \ac{hb} with index $\tau$, and then plug-in the results into \ac{hb} with index $\tau+1$ and perform another \ac{em} iteration. This process should continue until some convergence criteria are met, namely the change in the estimated parameters is small enough. In addition to the fact that the second alternative has lower computational complexity, it often yields much more stable results, as we observe in the empirical study (\secref{sec:EmpEval}).
%
%
\subsection{Inter Heartbeat Observation Noise Estimation} \label{sec:inter_obs_est}
To estimate $\eobscov{\tau, t}$, the inter-\ac{hb} observation noise covariance, we observe that $\ievolErr{\tau,t\given{T}}$, is the error covariance of the first stage estimation, namely
\begin{equation}
\hihb{\tau,t\given{\thor}}
-\ihb{\tau,t}\sim
\gnormal{0,
\ievolErr{\tau,t\given{T}}}.
\end{equation}
Since $\tilde{\gvec{y}}_{\tau,t}=\hihb{\tau,t\given{\thor}}$, the input for the second stage filtering is equal to the output of the first stage smoothing, then we can set $\ievolErr{\tau,t\given{T}}$ as $\Tilde{\mathcal{R}}_{\tau, t}$, an instantaneous estimate for the second-stage observation noise covariance.

To further smooth our estimate and to improve algorithm's stability, we exploit inter-\ac{hb} correlations by averaging the instantaneous estimates in a window of size $L=L_1+L_2+1$, and get
\begin{equation}
\hat{\mathcal{R}}_{\tau,t} = \frac{1}{L_1 + L_2+1} \sum_{\ell = -L_1}^{L_2} \Tilde{\mathcal{R}}_{\tau, t+\ell}.
\end{equation}
%
%
\subsection{Inter Heartbeat Process Noise Estimation - Single Channel} \label{sec:inter_evol_est}
To estimate the process covariance  $\eevolcov{\tau}$, we employ a \acl{mle} approach similar to that described in~\cite{de1988likelihood}. By examining the \ac{ss} model equation~\eqref{eq:ss_inter} and the \ac{kf} equation~\eqref{eq:kf1}, we establish that, for every $\forall t$, there exists a relationship between the innovation covariance $\eobsErr{\tau}$ and the process covariance $\eevolcov{\tau}$. An empirical estimate for $\eobsErr{\tau}$ is given by $\Delta\cdot\Delta^\top$, with $\Delta$ defined in~\eqref{eq:kf21}. In estimating the process covariance matrix, it is imperative to ensure the matrix is \ac{psd}. Consequently, we omit the correlation between different channels during the inter-filtering process and presume a diagonal structure for the covariance matrix. Namely:
\begin{equation}
\Tilde{\mathcal{Q}}_{\tau, t}=
\Tilde{\gvec{q}}^2_{\tau, t} \otimes \gvec{I}_{m\times m},
\end{equation}
where all diagonal entries are positive
\begin{equation}
\Tilde{\gvec{q}}^2_{\tau}\brackets{i}=
\max\set{\bar{\mathcal{Q}}_{\tau}\brackets{i,i},0}.
\end{equation}
Here, 
\begin{equation}
\bar{\mathcal{Q}}_{\tau}=
\Delta\cdot\Delta^\top
-
\heobscov{\tau}-
\eevolErr{\tau-1}
\end{equation}
To get a smoother and more robust estimator we can exploit both intra-\ac{hb} and inter-\ac{hb} correlations. We first 
average our estimator in a local time window
\begin{equation}
\mathcal{Q}_{\tau,t}^{*} =
\frac{1}{L_1 + L_2+1}\cdot\sum_{\ell = -L_1}^{L_2}\Tilde{\mathcal{Q}}_{\tau, t + \ell}  
\end{equation}
and then apply an exponential smoothing, i.e., a simple \ac{iir} filter, namely
\begin{equation}
\hat{\mathcal{Q}}_{\tau} = 
\alpha\cdot \mathcal{Q}_{\tau}^{*} + 
\brackets{1-\alpha}\cdot \hat{\mathcal{Q}}_{\tau-1}
\end{equation}
Where $0<\alpha <1$ is the forgetting factor.

While this option is simpler to compute, it is not always optimal as can be seen in the empirical results in \secref{sec:EmpEval}. Next, we relive that assumption, to exploit inter-channel correlation. 
%
%
%
%
%
\subsection{Inter Heartbeat Process Noise Estimation - Multi Channel}
To enhance our denoising algorithm's performance, we aim to relax the constraint that mandates the covariance matrix to be diagonal, thereby leveraging intra-heartbeat correlations. A primary requirement in estimating a covariance matrix is its \ac{psd}. While diagonal matrices with positive real entries are naturally \ac{psd}, this is not the case for arbitrary covariance matrices. Unlike diagonal matrices, there isn't a straightforward criterion to ensure a matrix's \ac{psd} property without significantly altering the matrix, such as modifying the eigenvalue matrix in a spectral decomposition. Thus, directly maximizing the closed-form solution may lead to filter instabilities. This is because the optimal estimate doesn't inherently adhere to the \ac{psd} condition, rendering it unsuitable for accurately defining the covariance of a Gaussian distribution. To address this challenge, especially in the context of our second estimation, it becomes imperative to explicitly enforce the \ac{psd} constraint. We propose accomplishing this using a variant of the \ac{rmgd} algorithm~\cite{bonnabel2013stochastic}. 

\ac{rmgd} is an advanced optimization technique that builds on the principles of traditional gradient descent. Its uniqueness lies in its ability to optimize on curved spaces called manifolds, making it especially valuable for matrices like the \ac{psd} ones of size ${m}\times{m}$. These matrices naturally span a manifold $\mathcal{S}^+$ within $\greal^{\frac{m\times\brackets{m+1}}{2}}$, visualized as a convex cone with symmetric matrices forming its tangent space, denoted 
$T_p \mathcal{S}^+=\mathcal{S}$~\cite{doi:10.1137/140978168}.

To optimize on $\mathcal{S}^+$, a tailored inner product on its tangent space, expressed as: 
\begin{equation}
g_p: T_p\boldsymbol{\cdot}\mathcal{S}^+ \rightarrow \mathbb{R},
\end{equation}
converts it into a Riemannian manifold~\cite{lee2018introduction}. This structured geometric space significantly impacts the gradient computations. Among the available metrics~\cite{bhatia2007positive, arsigny2007geometric}, we chose the log-Cholesky metric~\cite{doi:10.1137/18M1221084} due to its efficient closed-form solution for the exponential map, a crucial component of \ac{rmgd}.

The core \ac{rmgd} procedure begins by computing the gradient of our loss based on the current estimate. This gradient is then projected onto the manifold's tangent space. Using the exponential map, we traverse the negative gradient on $\mathcal{S}^+$, ensuring our estimate remains \ac{psd}.

The detailed mathematical description of this algorithm is out of the scope of this paper and it can be found in our code~\footnote{The source code used in our empirical study along with hyperparameters is at \url{https://github.com/KalmanNet/HKF_Thesis}.}

\subsection{Discussion}
The proposed \ac{hkf} leverages both intra- and inter-\ac{hb} relations through the introduced \ac{ss} model and can process multi-dimensional data. This feature enables the exploitation of spatial relations across \ac{ecg} signals recorded using multiple electrodes. Although our derivation involves several simplifications—particularly regarding the properties of process and observation noise covariances—these make the \ac{hkf} analytically tractable and ensure the computational demands remain relatively manageable. We observed improvements when integrating multi-channel information during the inter-filtering process. Furthermore, exploring additional techniques to achieve faster estimates of the covariance matrices could be fruitful.

A natural extension of \ac{hkf} would involve replacing the intra- and inter-HB smoothing and filtering with the recently proposed RTSNet and KalmanNet~\cite{RTSNet_TSP,KalmanNetTSP}, respectively. This modification would yield a hybrid \acl{mb} / \acl{dd} \ac{ecg} denoiser. We anticipate that such extensions, reserved for future exploration, will further boost performance by relaxing the aforementioned simplifications and enhancing \ac{hkf}'s noise tolerance.

Given that \ac{hkf} is anchored in a statistical Bayesian approach, it is possible to extend it into an algorithm that can filter \acp{ecg}, regardless of the presence of arrhythmia, and concurrently detect arrhythmias. This detection might employ a likelihood-ratio test contrasting the intra- and inter-filters. There are numerous potential applications for \ac{hkf}, such as its incorporation into smartwatches or its use in monitoring oxygen levels during labor. Each potential application presents its own set of challenges that warrant exploration. For instance, it would be vital to assess the interplay between the algorithms separating fetal from maternal signals in relation to \ac{hkf}, or to address the subtleties of lower amplitude signals typical in these scenarios.
%
%
\section{Empirical Study}\label{sec:EmpEval}
Next, we empirically evaluate the \ac{hkf} algorithm's performance. We begin by comparing \ac{hkf} with multiple benchmark algorithms, using two different \ac{ecg} recording \aclp{ds}. Initially, we discuss the performance of our algorithm for patients without arrhythmia, representing our algorithm's primary use case. We then showcase its efficacy for patients with arrhythmia and detail how the algorithm should be adapted for such scenarios. In the appendix, we provide an analysis illuminating the implications of certain design choices.

For the numerical evaluation, two distinct \ac{ecg} recording datasets were utilized. The first is the widely recognized \mitds dataset, which is publicly accessible~\cite{goldberger2000physiobank}, while the second dataset is proprietary. This selection ensures the algorithm's compatibility across various recording devices, sampling frequencies, and channel configurations.

Given that the datasets contain clean \ac{ecg} recordings, we artificially introduced \ac{ecg} to simulate a noisy environment. The \ac{mse} in relation to the clean signal served as our primary performance metric. However, it's important to acknowledge that in clinical evaluations, other factors, such as wave shape and clarity, are also significant.
%
%
\subsection{Benchmark Algorithms}
For our evaluation, the complete \ac{hkf} was employed. The intermediary output of the first stage \ac{ks} operation carries the label (\ac{ks}-intra), and the final independent \ac{hkf} output is denoted as (\ac{hkf}). We've also presented the multichannel estimate under the label (\ac{hkf} \ac{rmgd}). Furthermore, results from applying only the inter-\ac{hb} filtering, as highlighted in~\cite{VullingsVB11}, are labeled as (\ac{kf}-inter).

In addition to these, we included a filtered estimate generated by a data-driven convolutional \ac{ae} following the methodologies in~\cite{8693790, Fotiadou2020MultiChannelFE}, marked as (\ac{ae}). To ensure optimal \ac{ae} performance, training encompassed the entire dataset, excluding the tested subject. The full details are provided in our code. 
%
%
%
%
\subsection{Patients without Arrhythmia - Proprietary Dataset}
\begin{figure*}[]
\begin{subfigure}[a]{0.5\columnwidth}
\centering
\includegraphics[width=1.1\columnwidth]{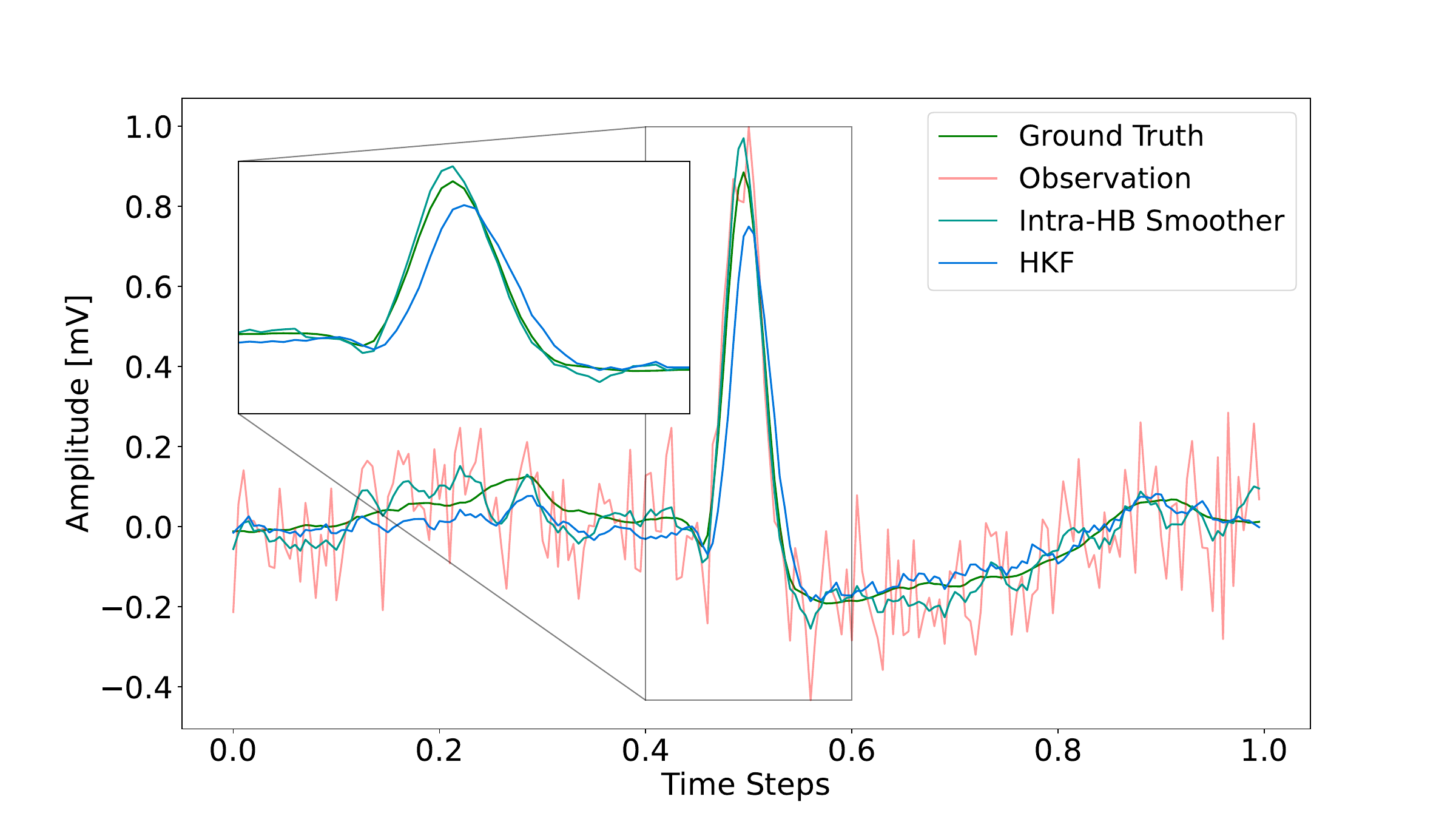}
\end{subfigure}
\begin{subfigure}[a]{0.5\columnwidth}
\centering
\includegraphics[width=1.1\columnwidth]
{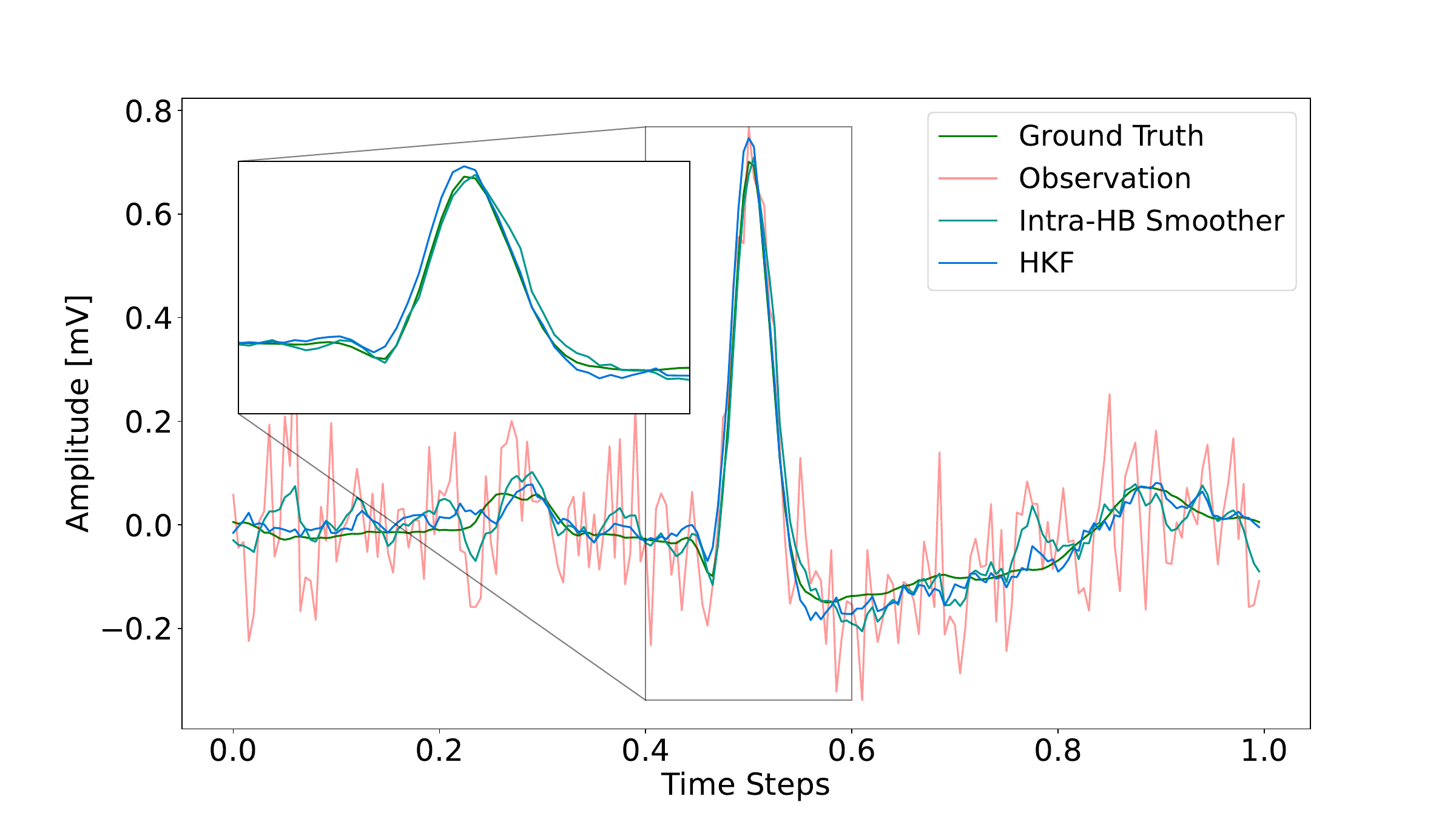}
\end{subfigure}
\begin{subfigure}[a]{0.5\columnwidth}
\centering
\includegraphics[width=1.1\columnwidth]
{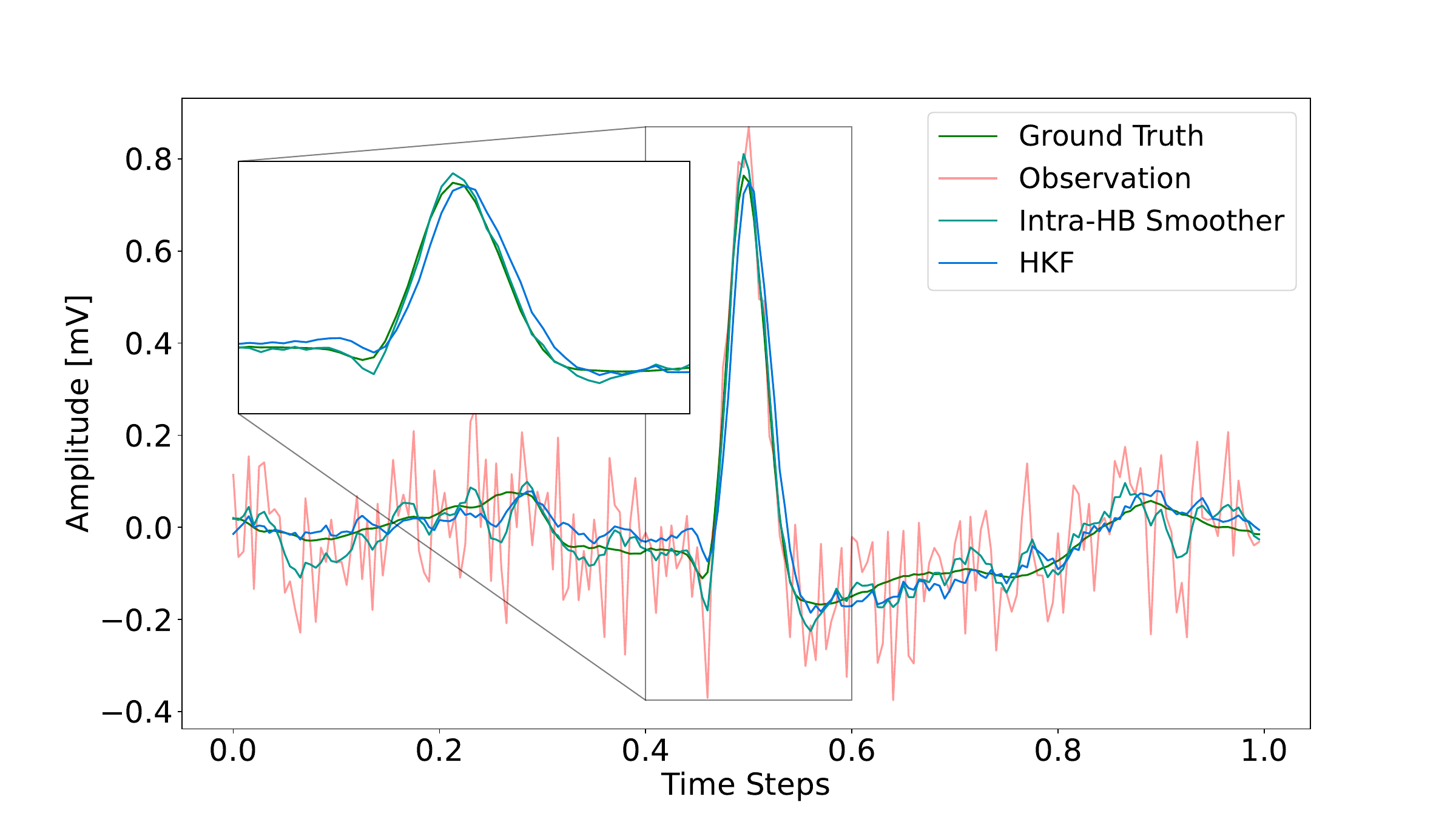}
\end{subfigure}
\begin{subfigure}[a]{0.5\columnwidth}
\centering
\includegraphics[width=1.1\columnwidth]
{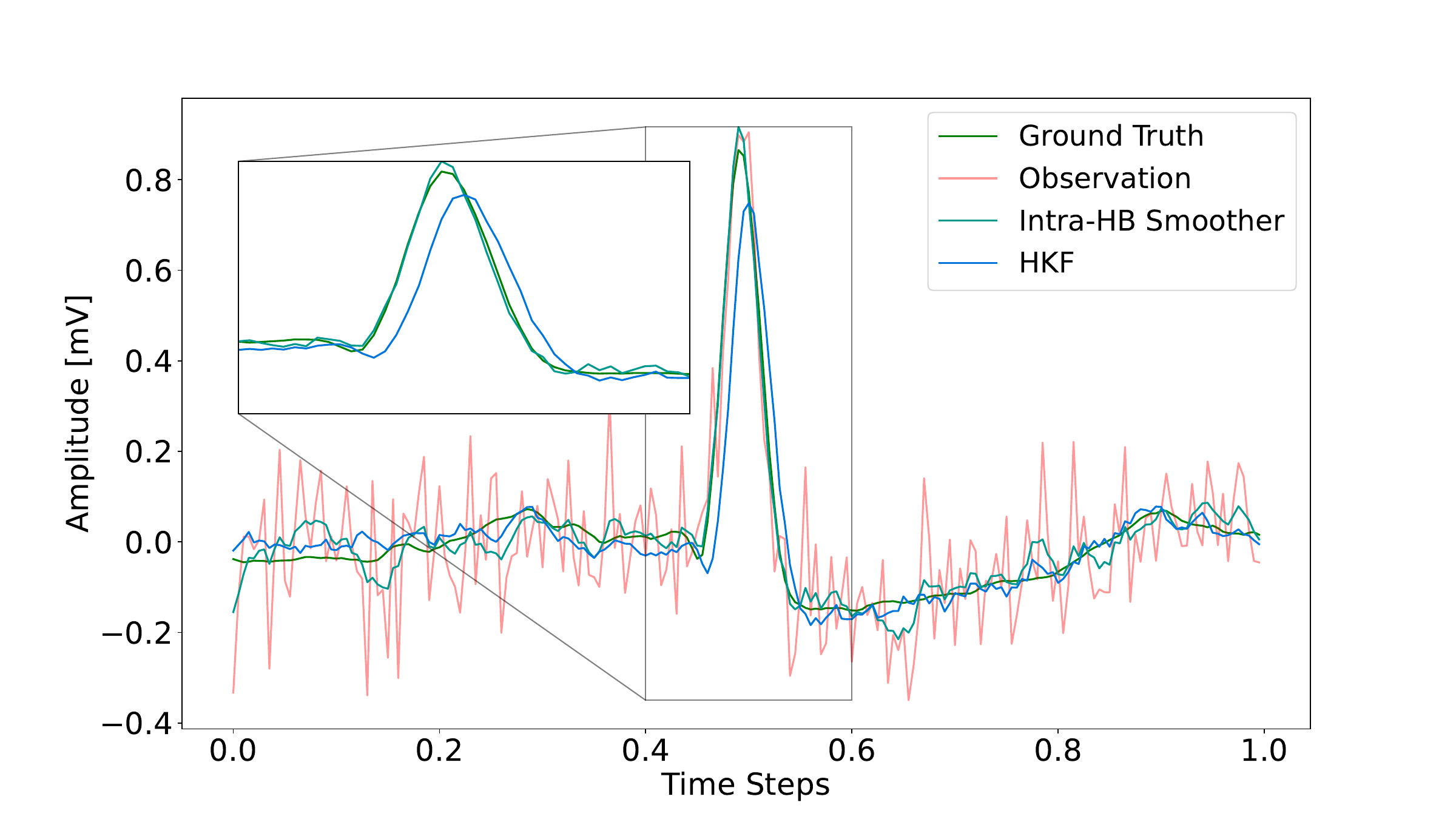}
\end{subfigure}
\caption{Single \ac{hb} - Patient 0 - Proprietary}
\label{fig:rik_pat_0_single}
%
%
%
\begin{subfigure}[a]{0.5\columnwidth}
\centering
\includegraphics[width=1.1\columnwidth]{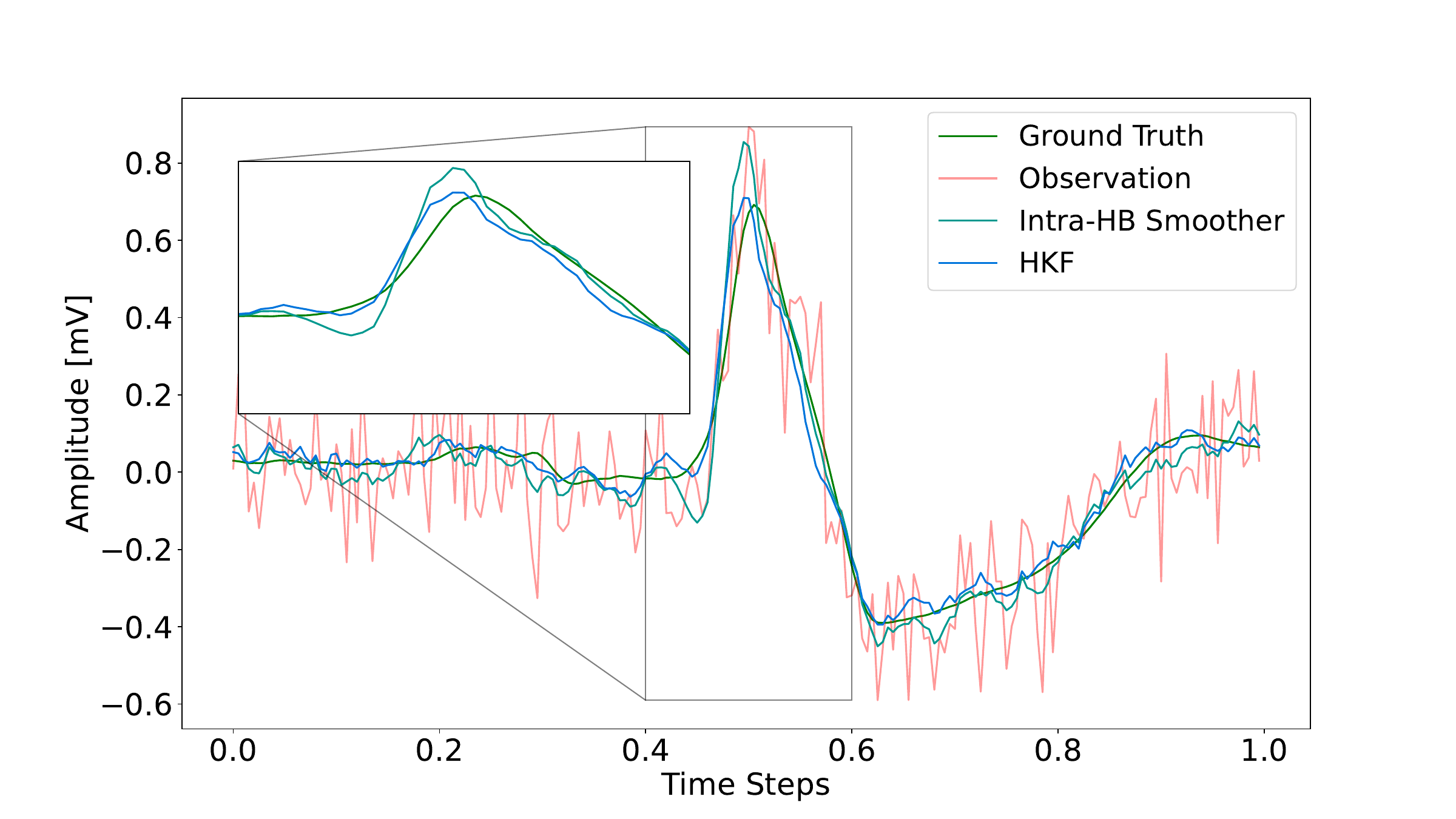}
\end{subfigure}
\begin{subfigure}[a]{0.5\columnwidth}
\centering
\includegraphics[width=1.1\columnwidth]
{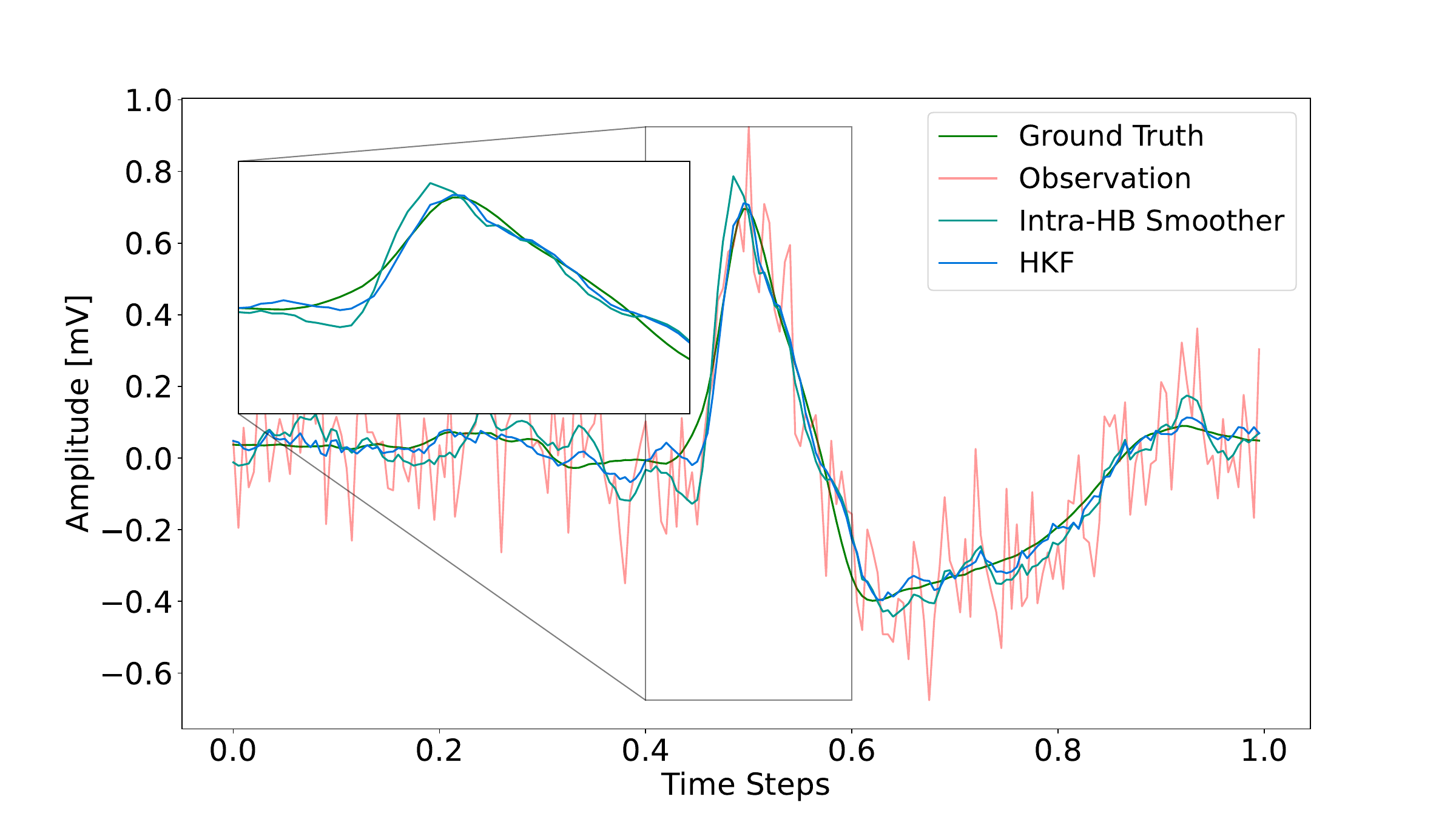}
\end{subfigure}
\begin{subfigure}[a]{0.5\columnwidth}
\centering
\includegraphics[width=1.1\columnwidth]
{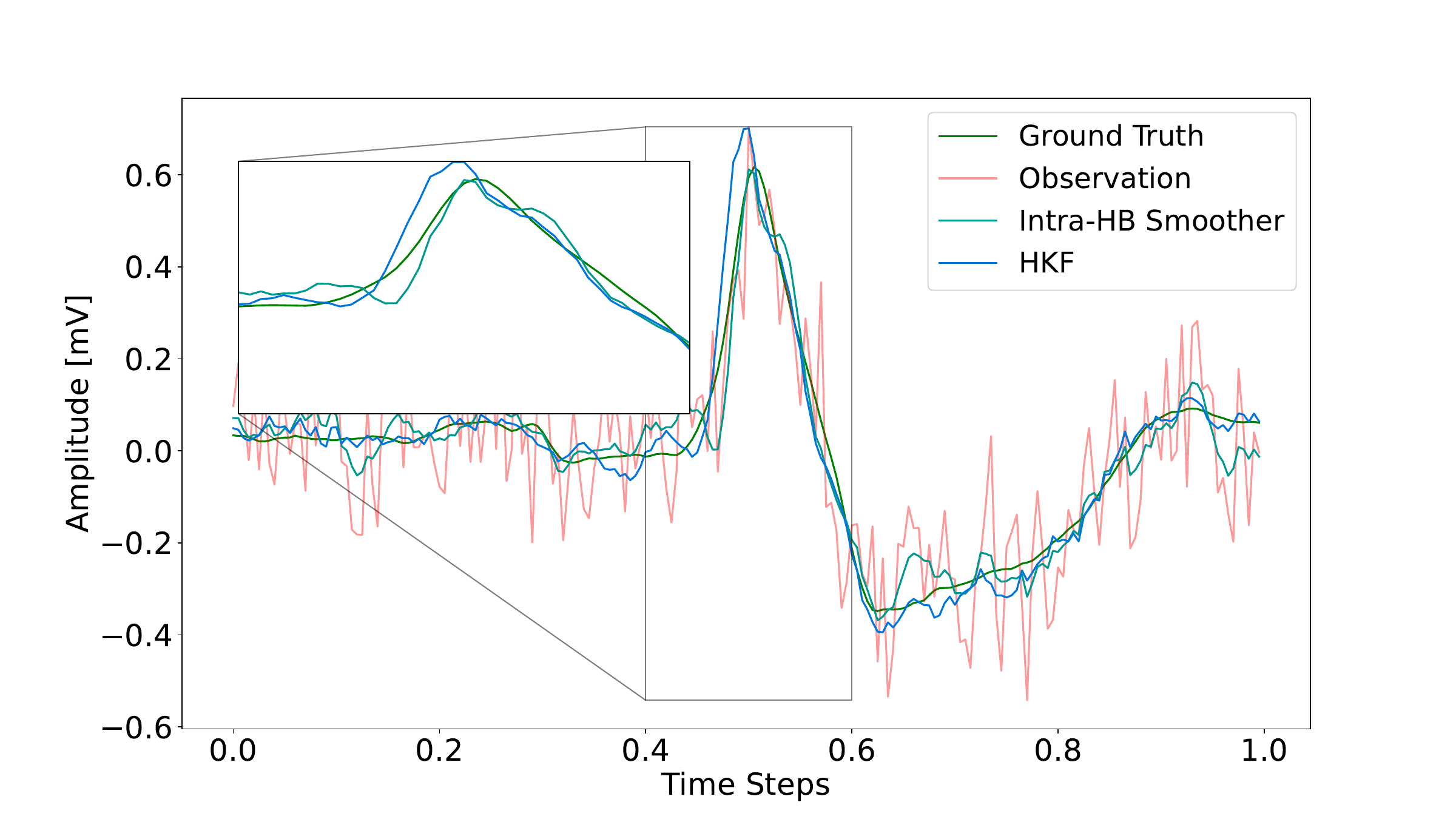}
\end{subfigure}
\begin{subfigure}[a]{0.5\columnwidth}
\centering
\includegraphics[width=1.1\columnwidth]
{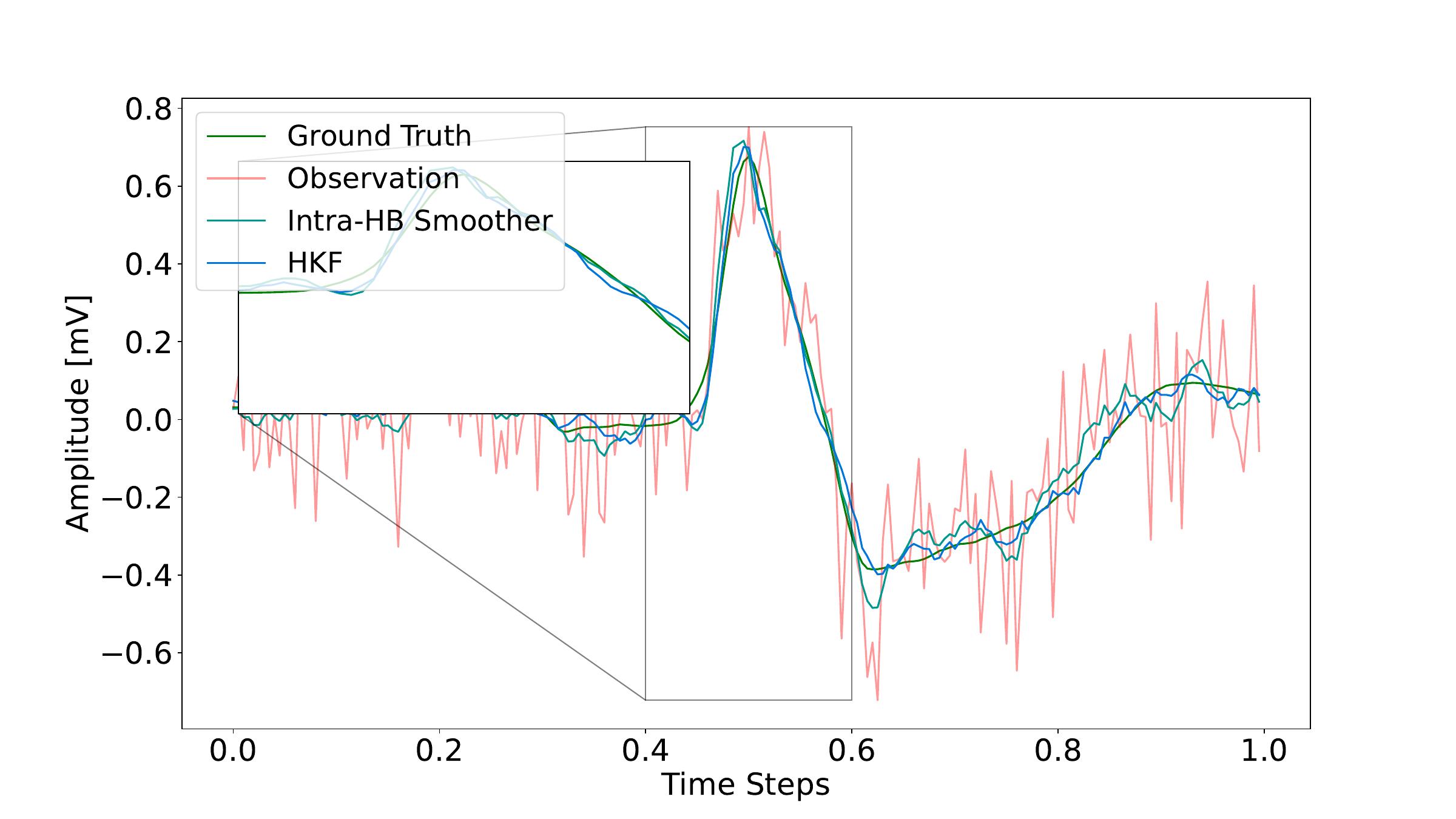}
\end{subfigure}
\caption{Single \ac{hb}  for patient 1 - Proprietary}
\label{fig:rik_pat_1_single}
%
%
%
\begin{subfigure}[a]{0.5\columnwidth}
\centering
\includegraphics[width=1.1\columnwidth]{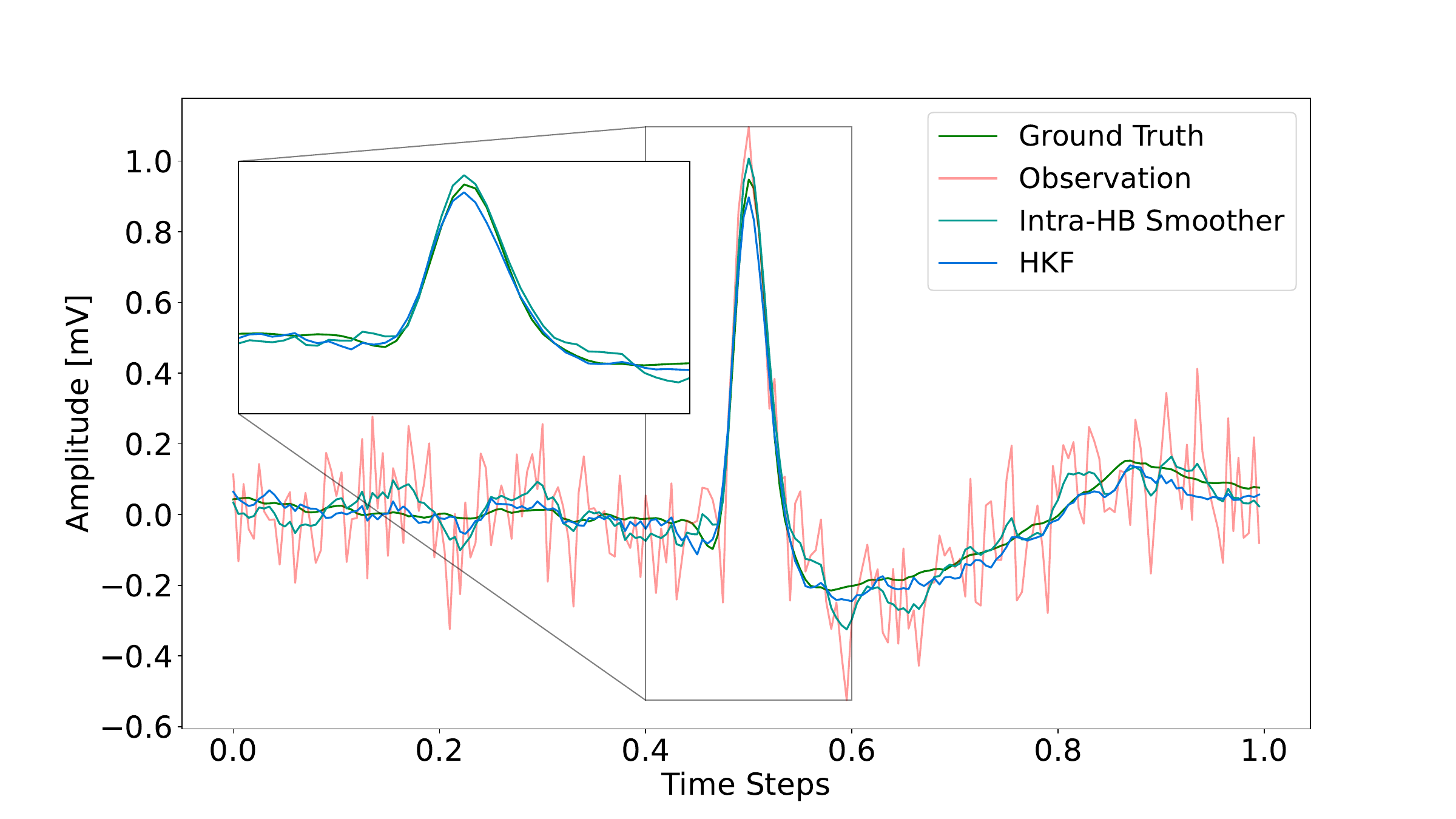}
\end{subfigure}
\begin{subfigure}[a]{0.5\columnwidth}
\centering
\includegraphics[width=1.1\columnwidth]
{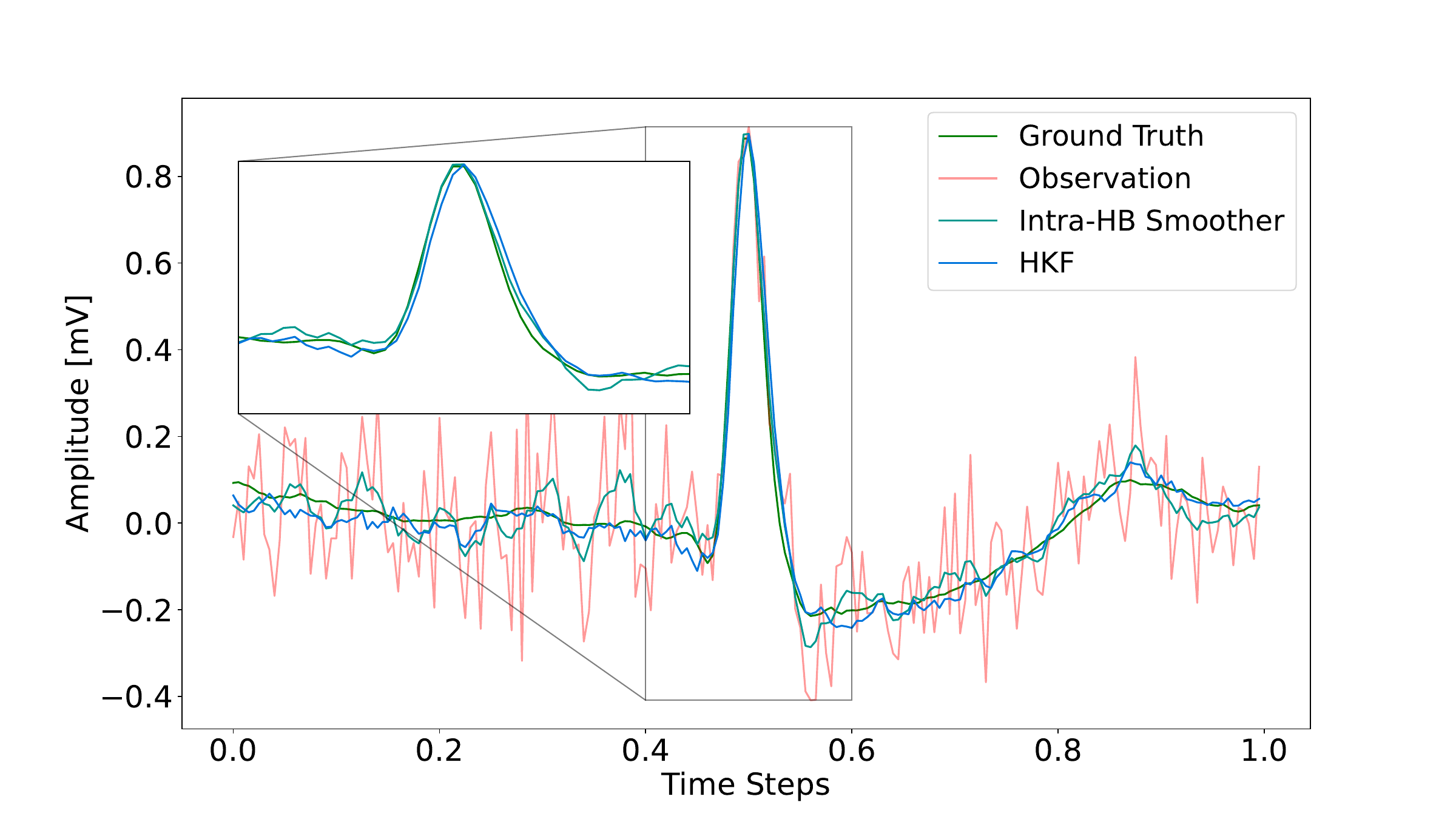}
\end{subfigure}
\begin{subfigure}[a]{0.5\columnwidth}
\centering
\includegraphics[width=1.1\columnwidth]
{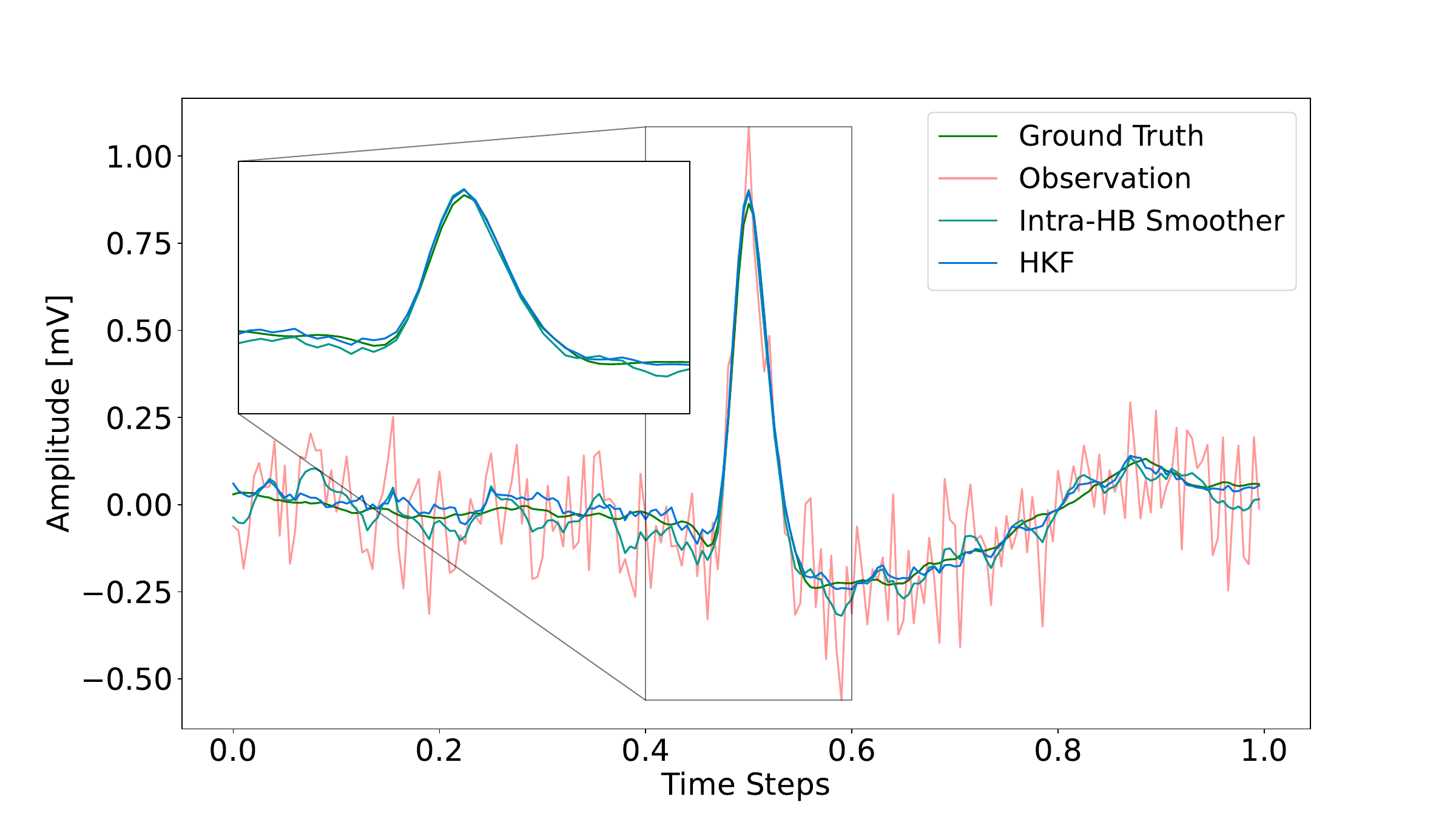}
\end{subfigure}
\begin{subfigure}[a]{0.5\columnwidth}
\centering
\includegraphics[width=1.1\columnwidth]
{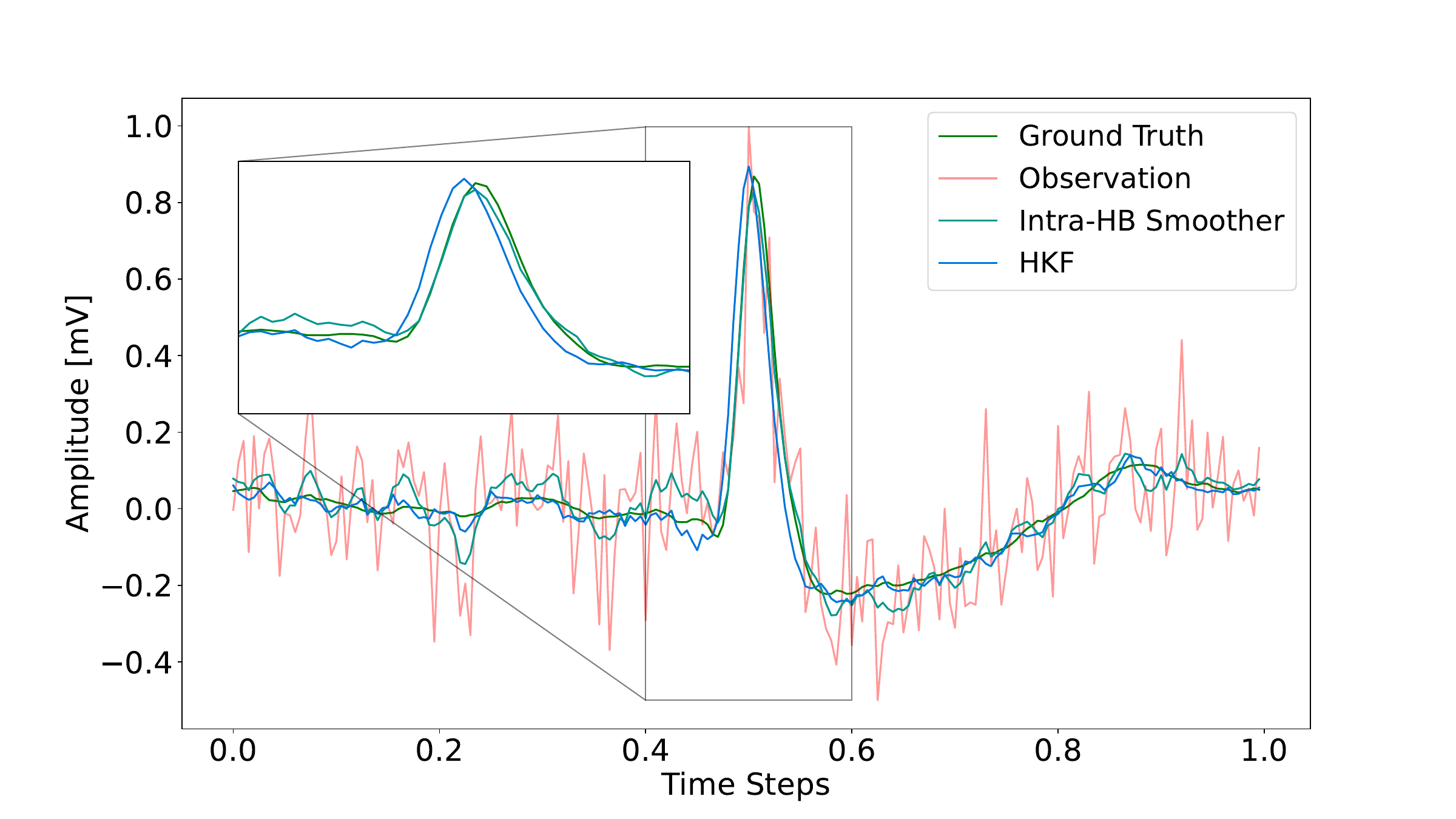}
\end{subfigure}
\caption{Single \ac{hb} - Patient 2 - Proprietary}
\label{fig:rik_pat_2_single}
%
%
%
\begin{subfigure}[a]{0.5\columnwidth}
\centering
\includegraphics[width=1.1\columnwidth]{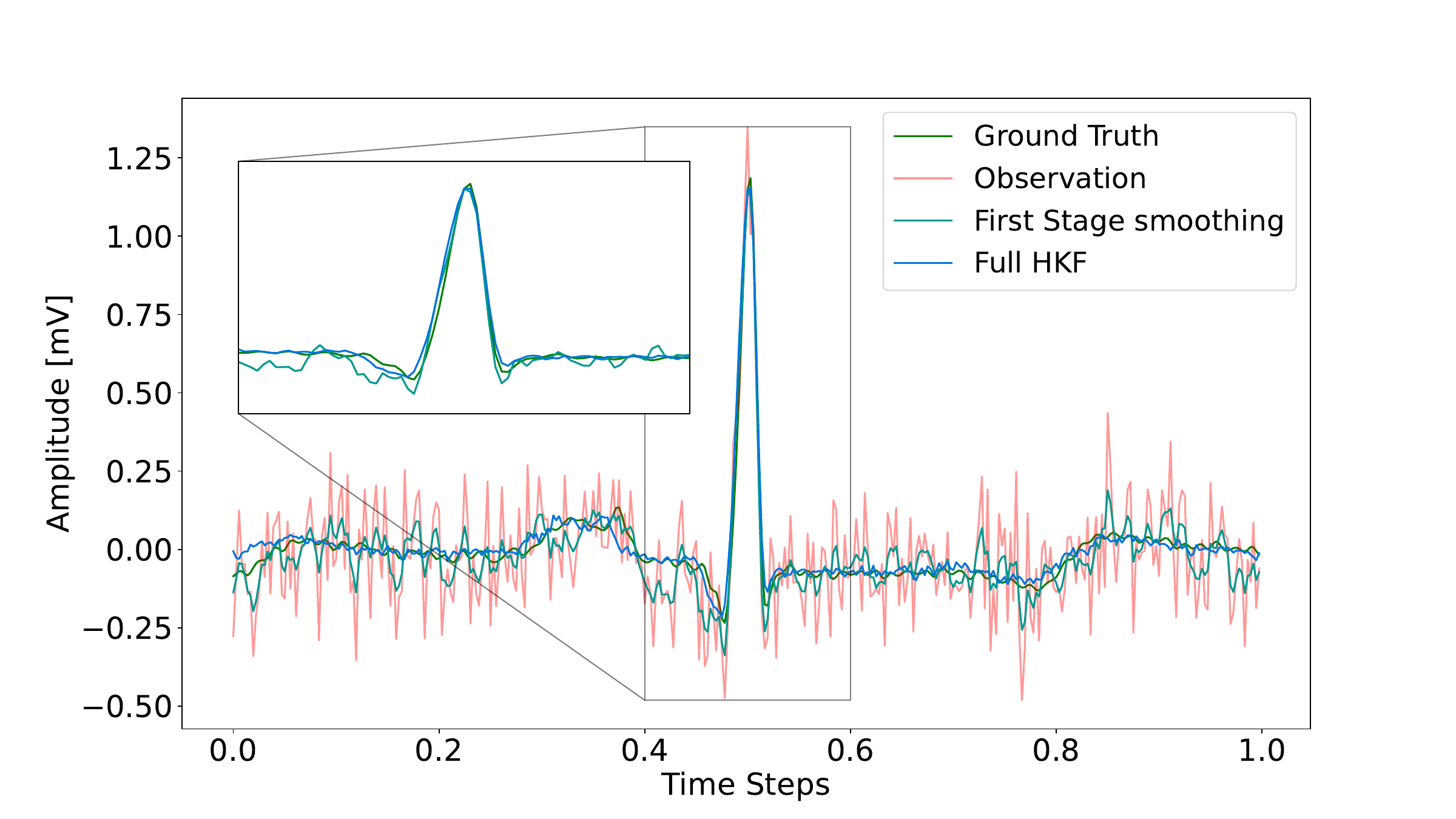}
\end{subfigure}
\begin{subfigure}[a]{0.5\columnwidth}
\centering
\includegraphics[width=1.1\columnwidth]
{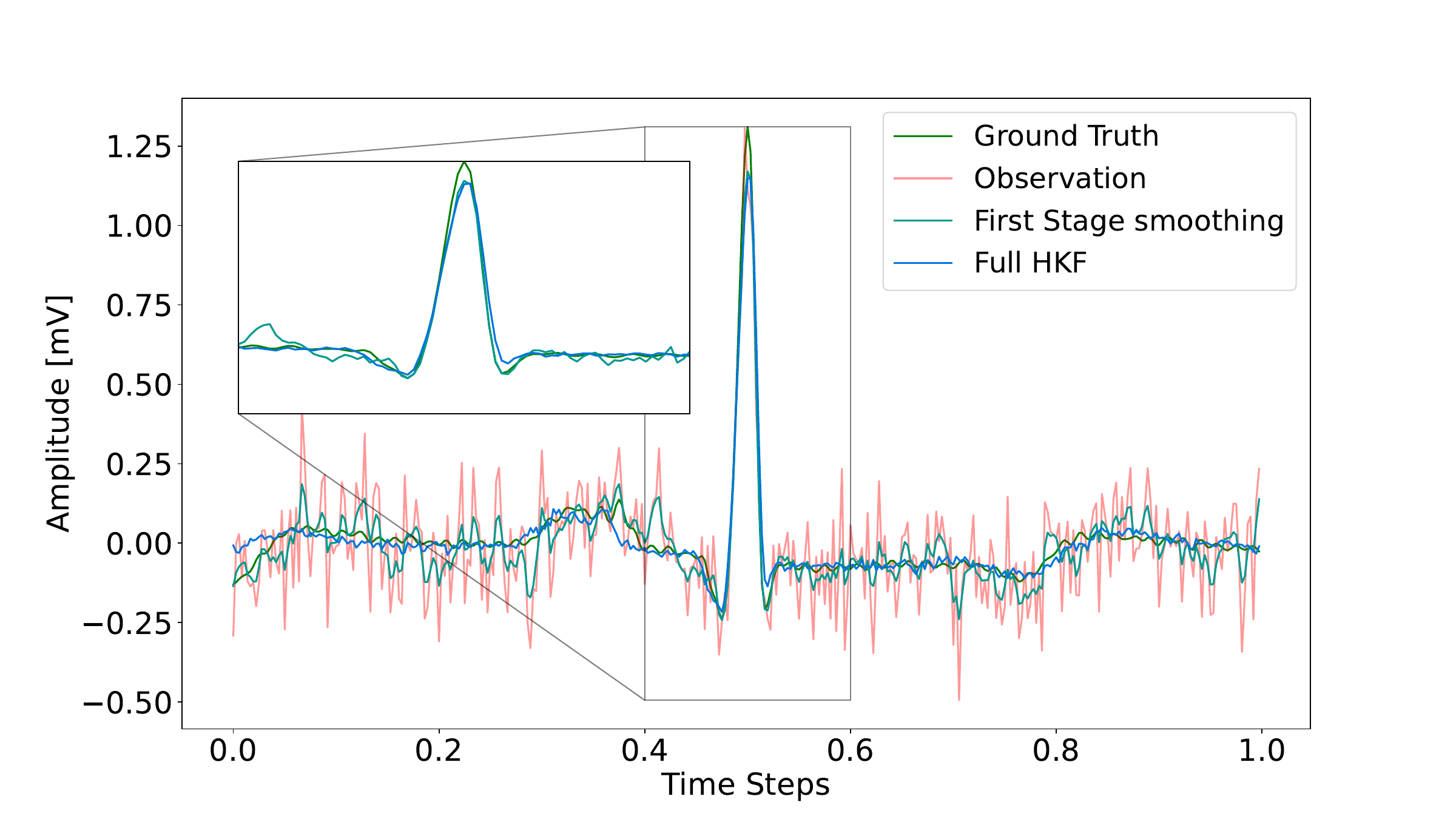}
\end{subfigure}
\begin{subfigure}[a]{0.5\columnwidth}
\centering
\includegraphics[width=1.1\columnwidth]
{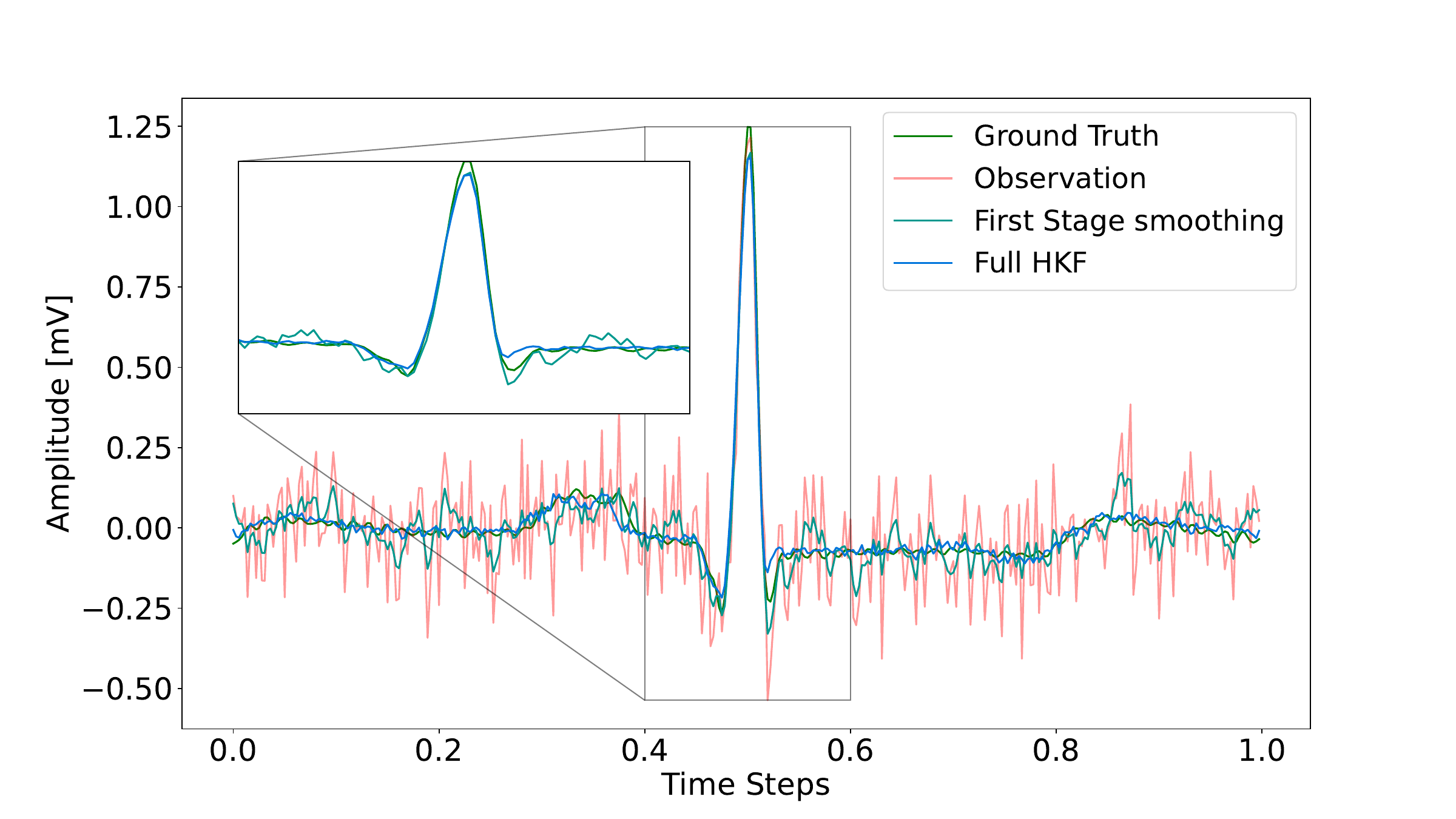}
\end{subfigure}
\begin{subfigure}[a]{0.5\columnwidth}
\centering
\includegraphics[width=1.1\columnwidth]
{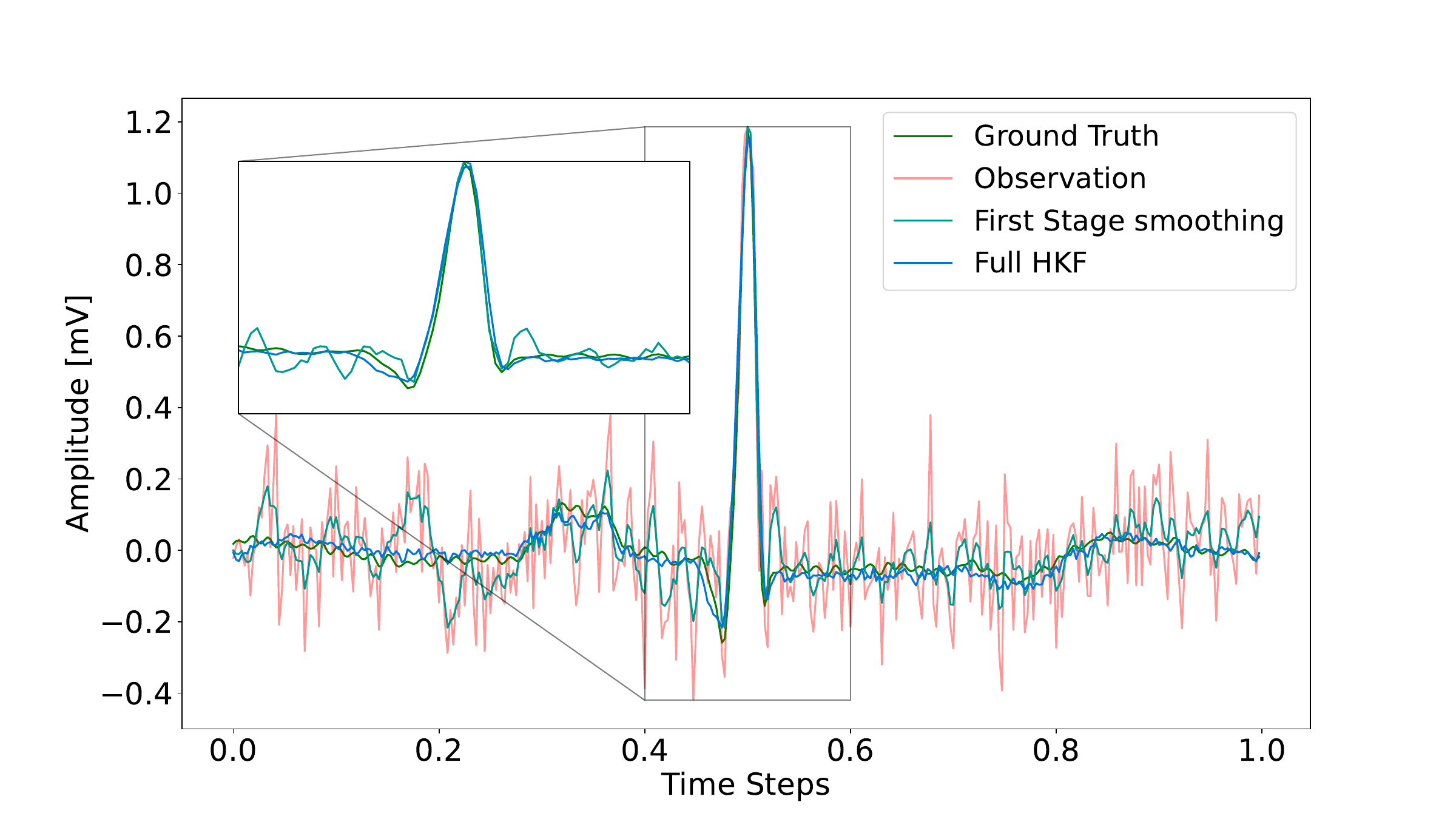}
\end{subfigure}
\caption{Single \ac{hb} - Patient 100 - MIT-BIH}
\label{fig:mit_pat_0_single}
%
%
%
\begin{subfigure}[a]{0.5\columnwidth}
\centering
\includegraphics[width=1.1\columnwidth]{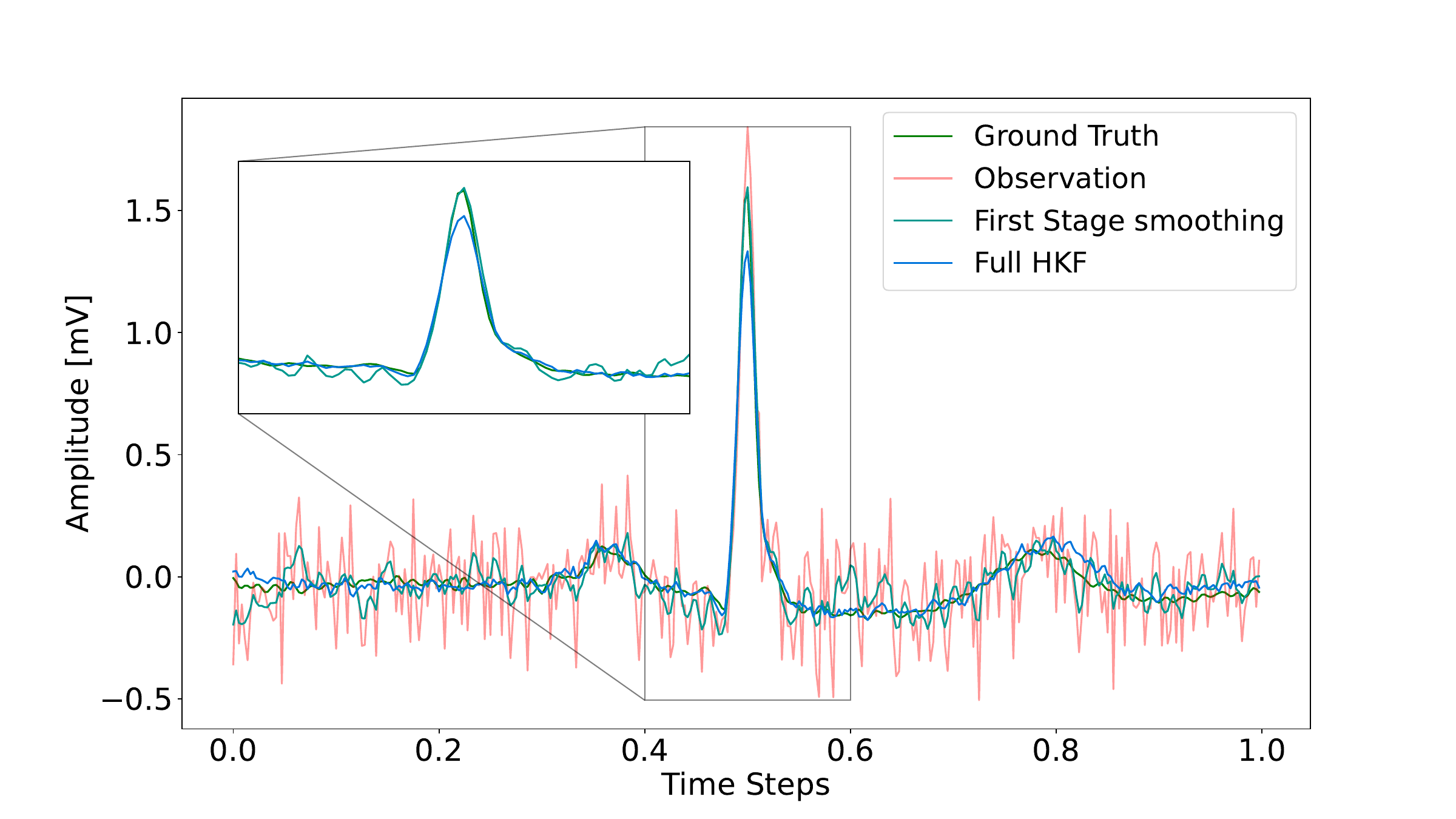}
\end{subfigure}
\begin{subfigure}[a]{0.5\columnwidth}
\centering
\includegraphics[width=1.1\columnwidth]
{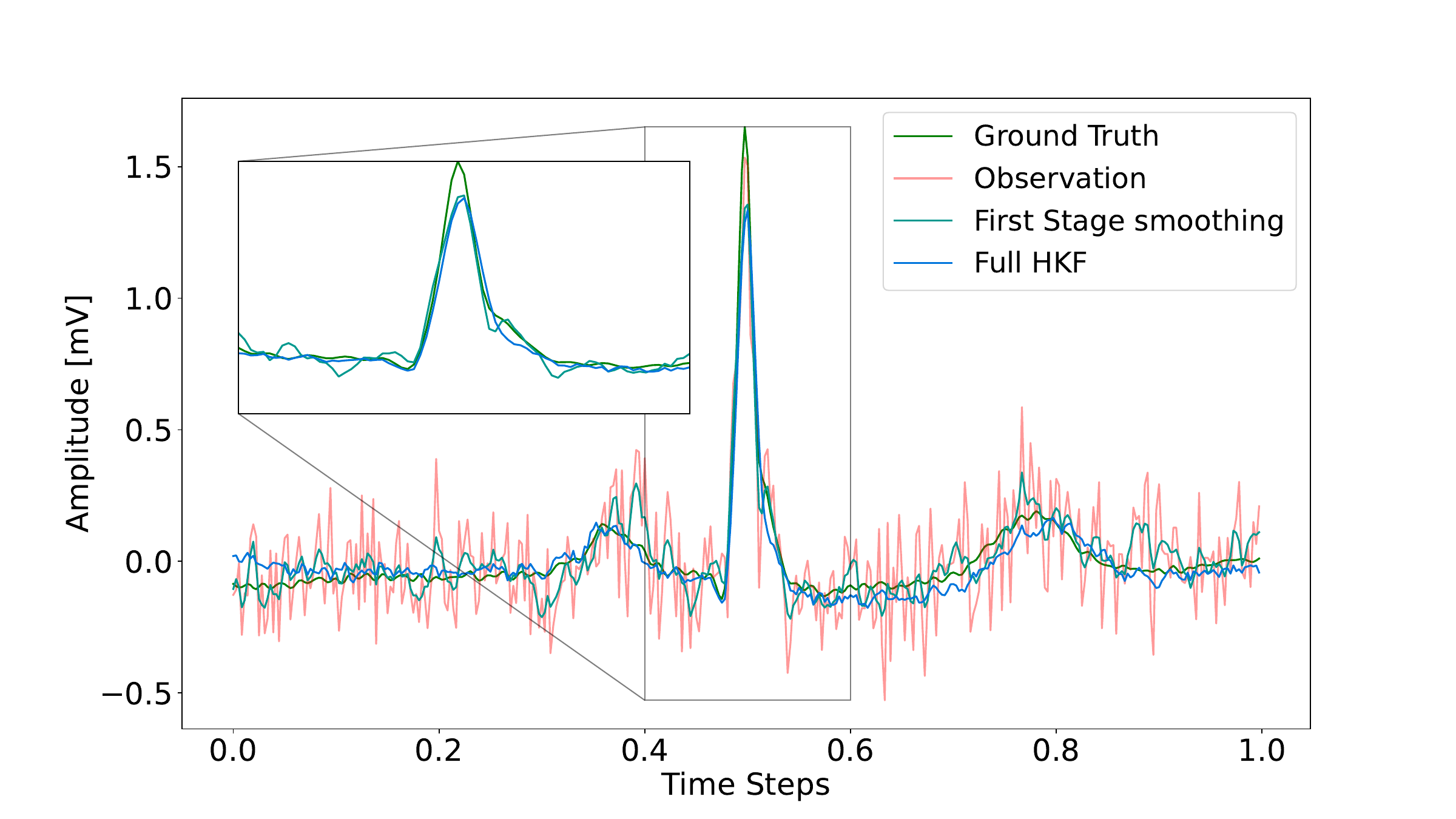}
\end{subfigure}
\begin{subfigure}[a]{0.5\columnwidth}
\centering
\includegraphics[width=1.1\columnwidth]
{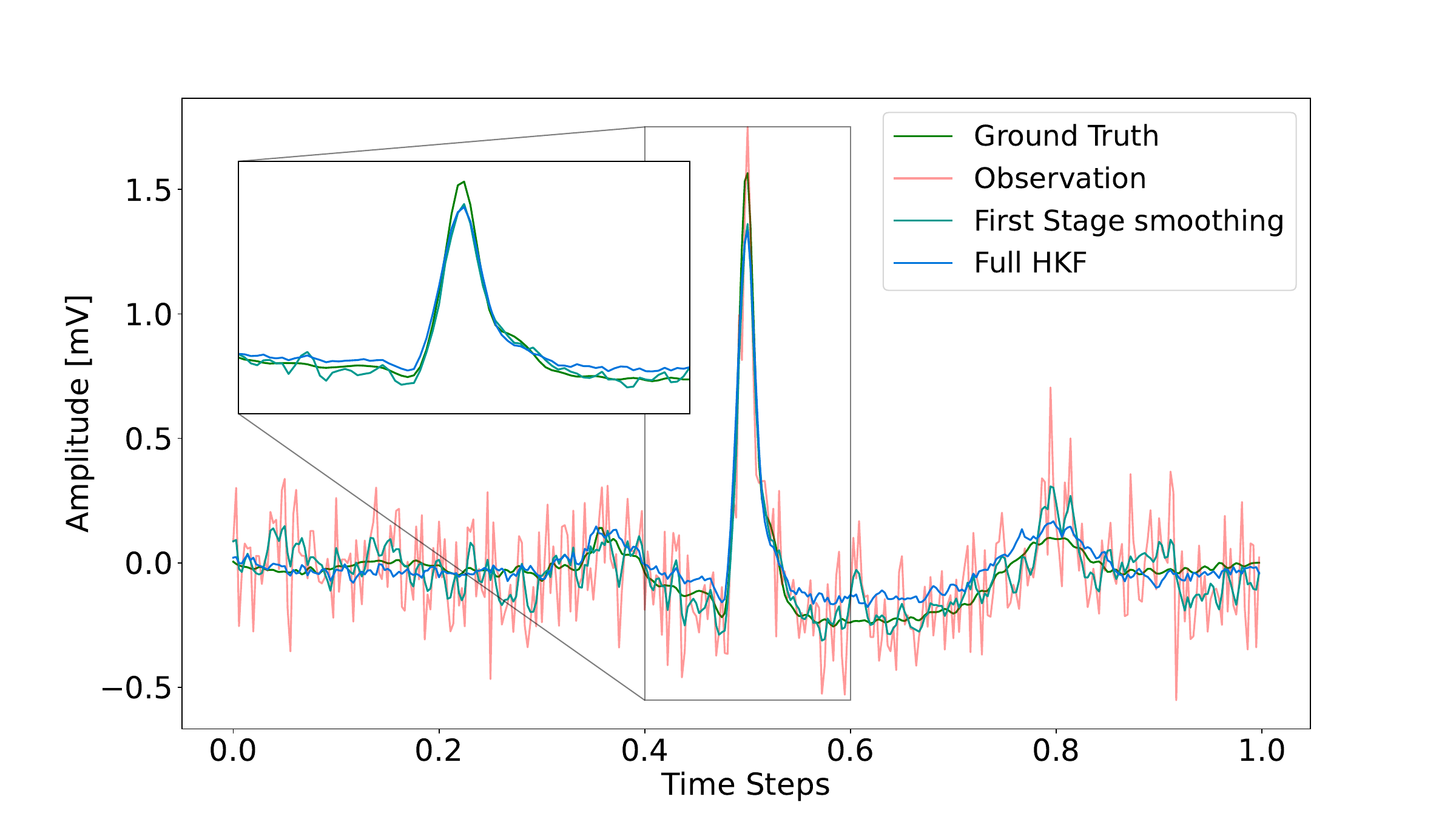}
\end{subfigure}
\begin{subfigure}[a]{0.5\columnwidth}
\centering
\includegraphics[width=1.1\columnwidth]
{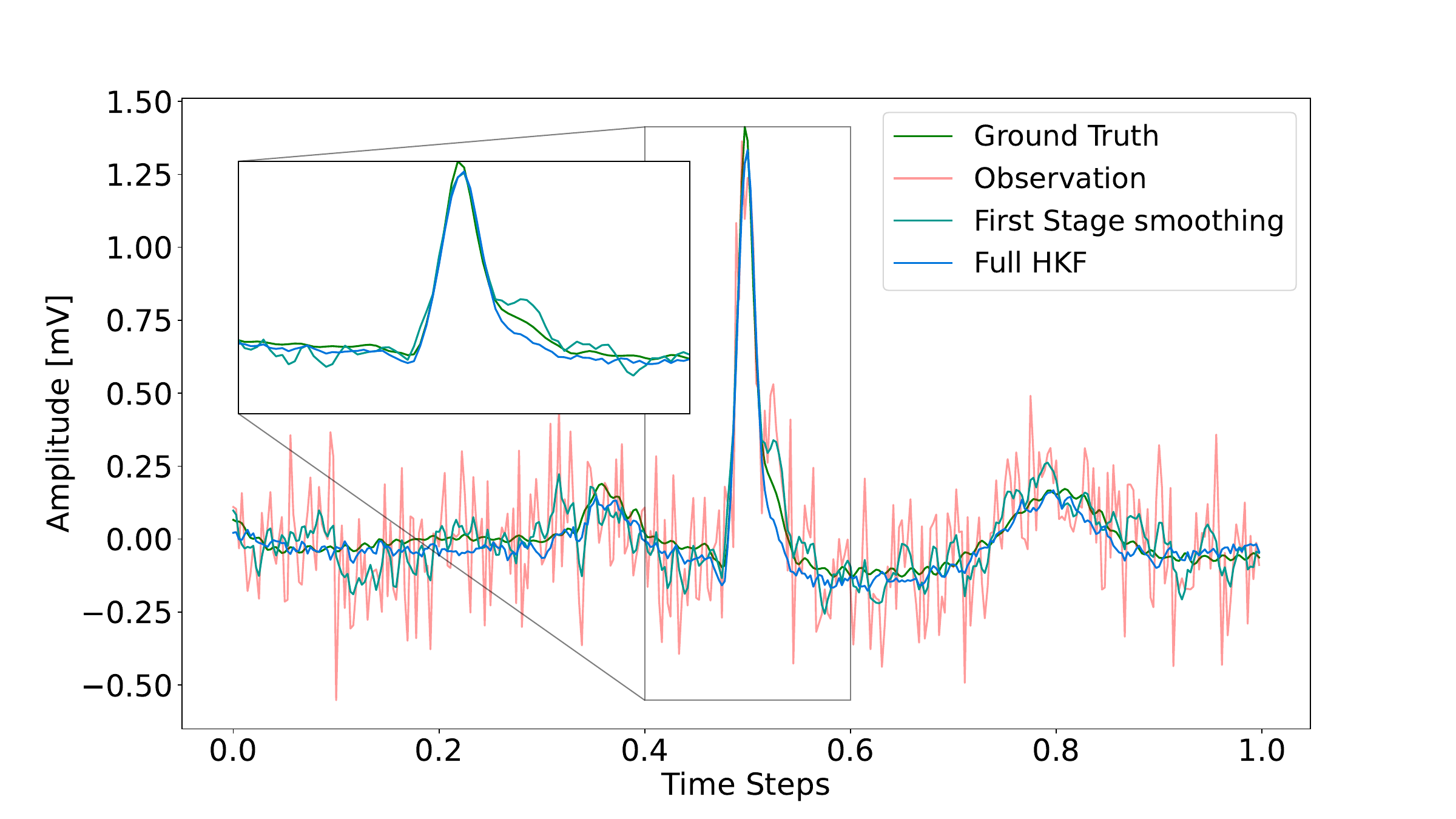}
\end{subfigure}
\caption{Single \ac{hb} - Patient 101 - MIT-BIH}
\label{fig:mit_pat_1_single}
\end{figure*}

We next turn our attention to the evaluation of our algorithm on patients without arrhythmia, using a proprietary dataset. This dataset offers clean \ac{ecg} recordings of adults, captured at a sampling rate of $500\si{\hertz}$. To emulate fetal $\ac{ecg}$ signals, two modifications were applied to this data:
\begin{enumerate}
\item Given that the fetal \ac{hr} is typically double that of an adult, the \ac{ecg} signals were subsampled by a factor of two. This creates signals that appear to have a \ac{hr} twice as high, with a resultant sampling rate of $250\si{\hertz}$.
\item An attenuation of the signal amplitude was performed to simulate the often weaker amplitude seen in fetal \acp{ecg}. This diminished amplitude can be attributed to the smaller size of the fetal heart and the typical recording conditions where the \ac{ecg} is taken from the mother's abdomen rather than directly from the fetal thorax.
\end{enumerate}
Our algorithm's robust performance, especially when compared with alternative methods on \ac{ecg} signals corrupted to $0\dB$ \ac{snr}, is detailed in \tbref{tbl:prop}. Given that the \ac{ecg} signal was captured using $m = 12$ observation channels, our \ac{hkf} with \ac{rmgd} effectively leverages correlations between these multiple channels, consistently outpacing other methods in performance. Visual representations showcasing our algorithm's efficacy can be viewed in the subsequent figures: \ref{fig:rik_pat_0_single}, \ref{fig:rik_pat_0}, \ref{fig:rik_pat_1_single}, \ref{fig:rik_pat_1}, \ref{fig:rik_pat_2_single}, and \ref{fig:rik_pat_2}.

%
%
\subsection{Patients without Arrhythmia - MIT BIH Dataset}
Next, we assess our algorithm using patients without arrhythmia from the \mitds (Arrhythmia) \acl{ds}. This \acl{ds} comprises $48$ two-channel ambulatory \ac{ecg} signals recorded by the BIH Arrhythmia Laboratory between 1975 and 1979. These signals were recorded at a rate of $360$ samples per second with an $11$-bit resolution over a $10\unit{\milli\volt}$ range. This \acl{ds} is widely recognized in the \ac{ecg} denoising community and serves as a reliable benchmark for evaluating algorithm performance. Examples of its application can be found in references~\cite{weng2006ecg,sayadi2006ecg, SinghP21}.

\begin{figure*}[]
\centering
%
\begin{subfigure}[a]{0.65\columnwidth}
\centering
\includegraphics[width = 1\columnwidth]{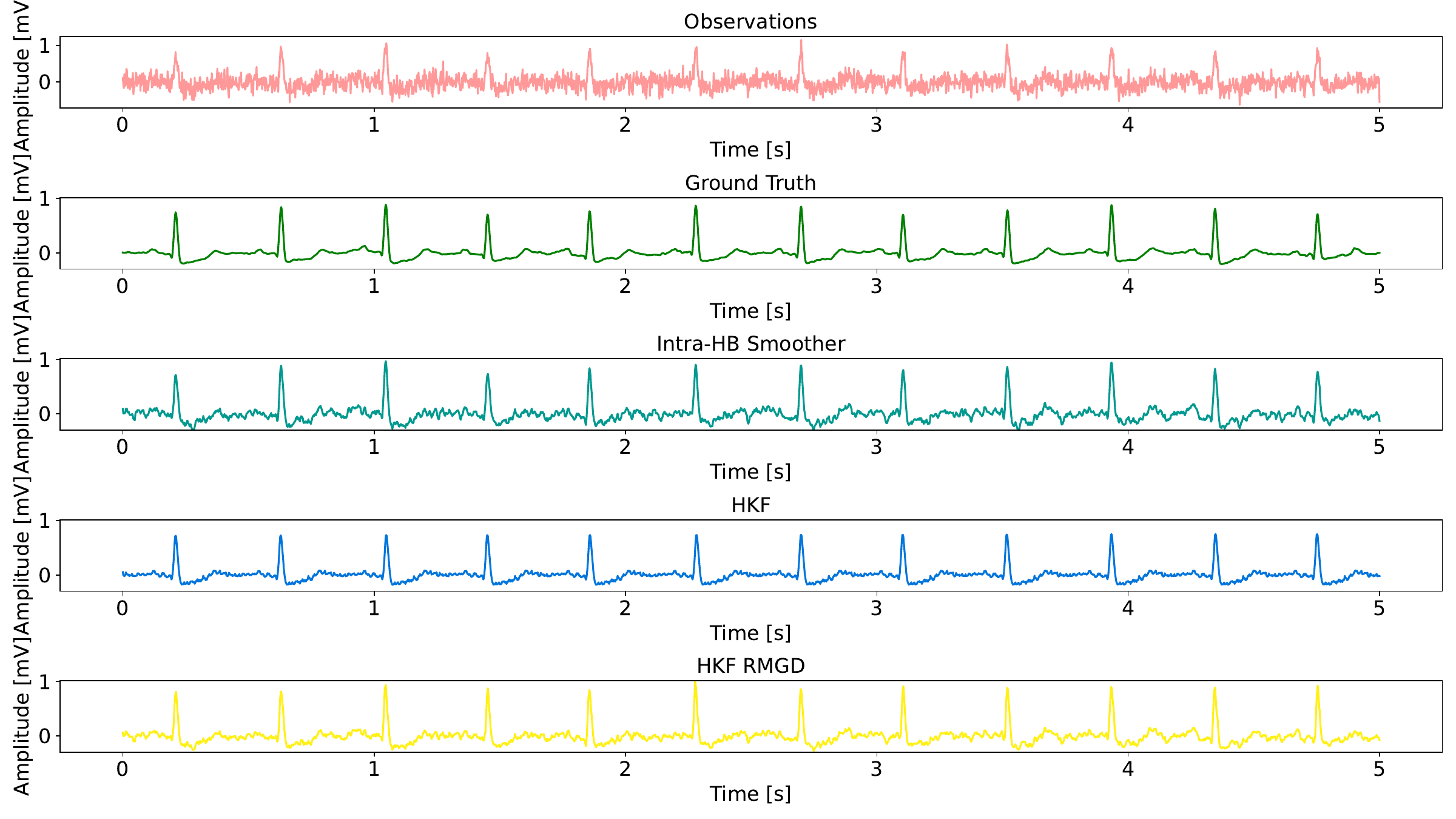}
\caption{Patient 0 - Proprietary}
\label{fig:rik_pat_0}  
\end{subfigure}
%
%
\begin{subfigure}[a]{0.65\columnwidth}
\centering
\includegraphics[width = 1\columnwidth]{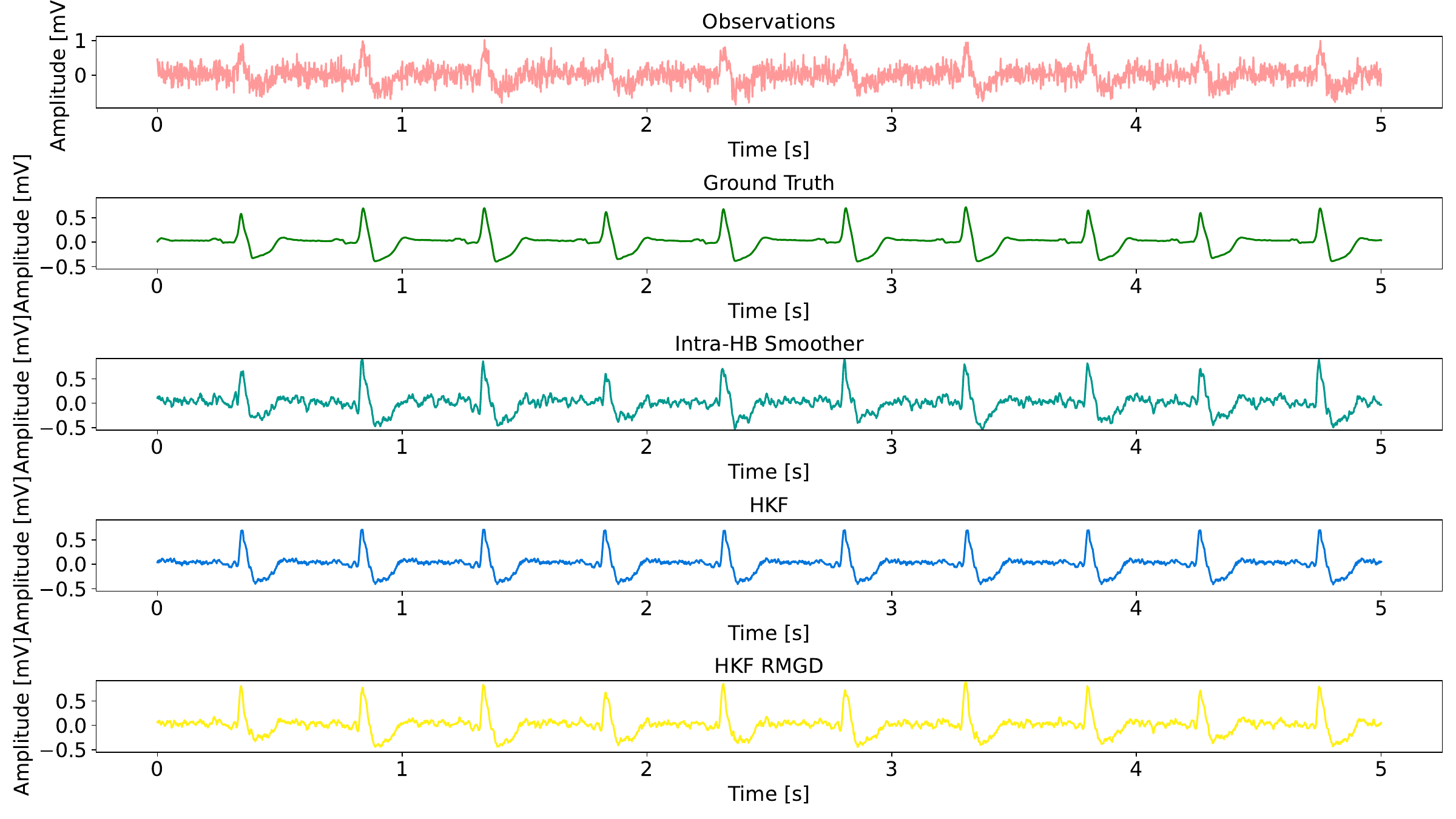}
\caption{Patient 1 - Proprietary}
\label{fig:rik_pat_1}  
\end{subfigure}
%
%
\begin{subfigure}[a]{0.65\columnwidth}
\centering
\includegraphics[width = 1\columnwidth]{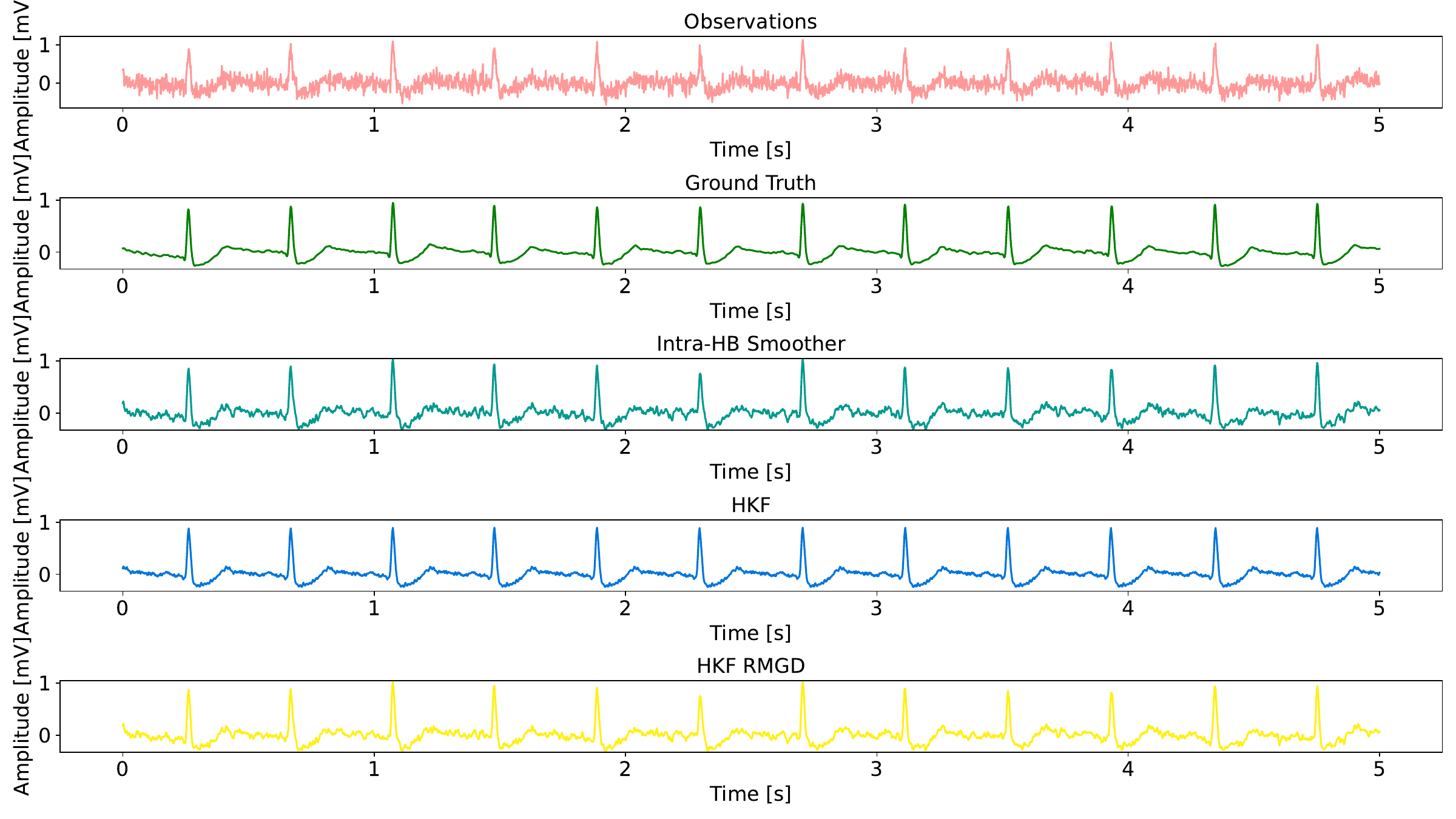}
\caption{Patient 2 - Proprietary}
\label{fig:rik_pat_2}  
\end{subfigure}
%
%
\centering
%
\begin{subfigure}[a]{0.49\columnwidth}
\centering
\includegraphics[width = 1\columnwidth]{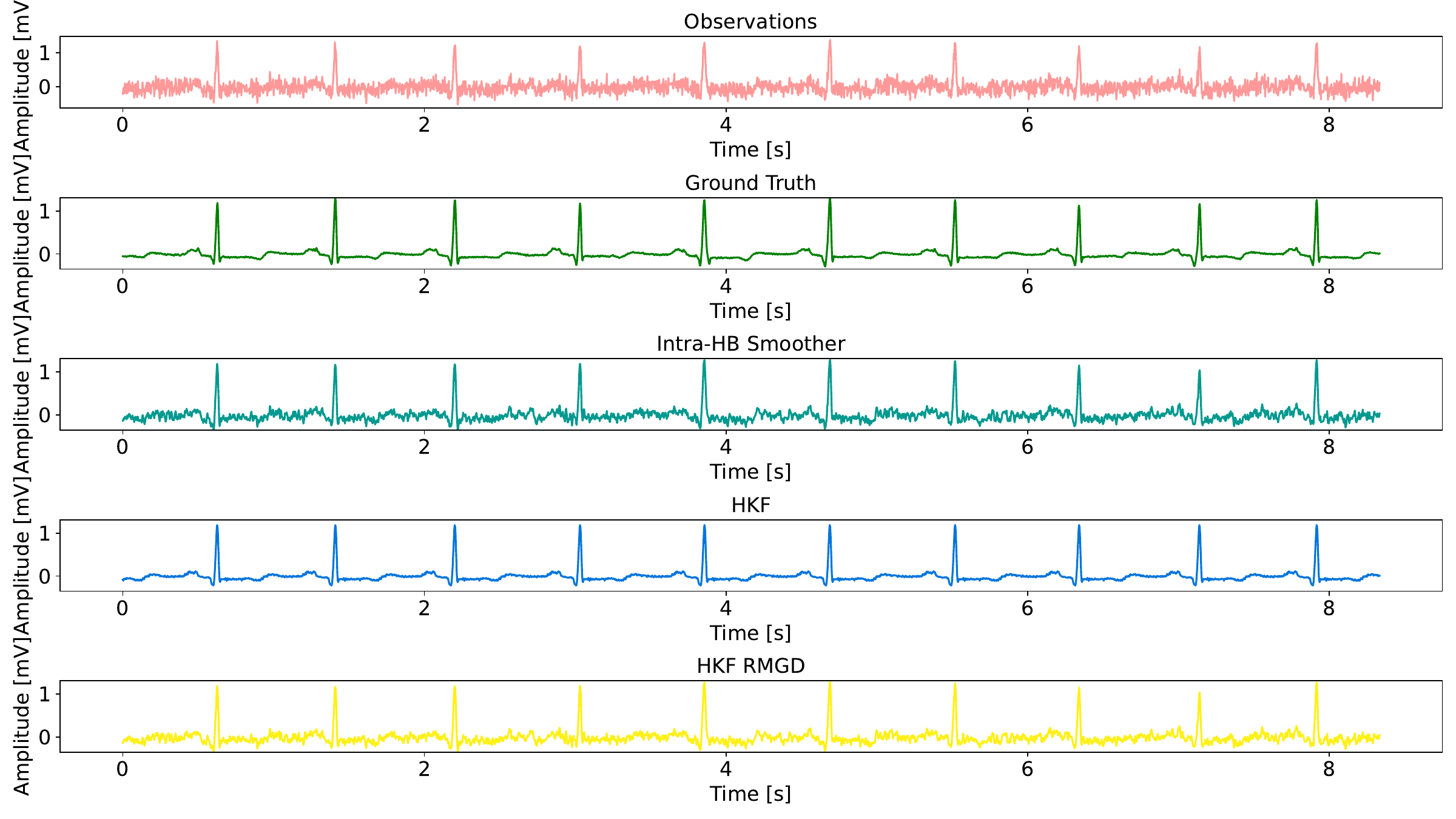}
\caption{Patient 100 - MIT-BIH}
\label{fig:mit_pat_0}  
\end{subfigure}
%
%
\begin{subfigure}[a]{0.49\columnwidth}
\centering
\includegraphics[width = 1\columnwidth]{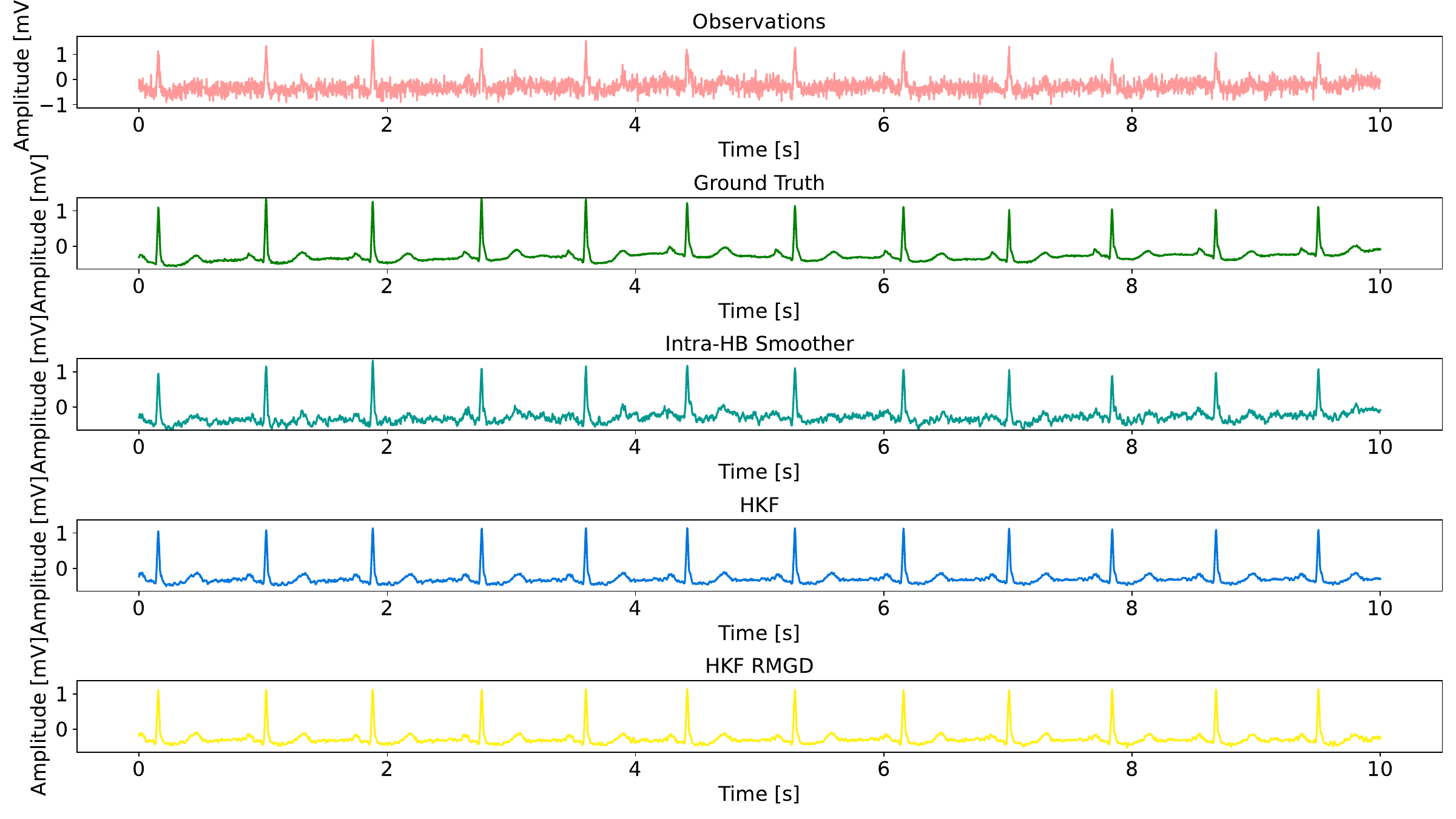}
\caption{Patient 101 - MIT-BIH}
\label{fig:mit_pat_1}  
\end{subfigure}
%
%
\begin{subfigure}[a]{0.49\columnwidth}
\centering
\includegraphics[width = 1\columnwidth]{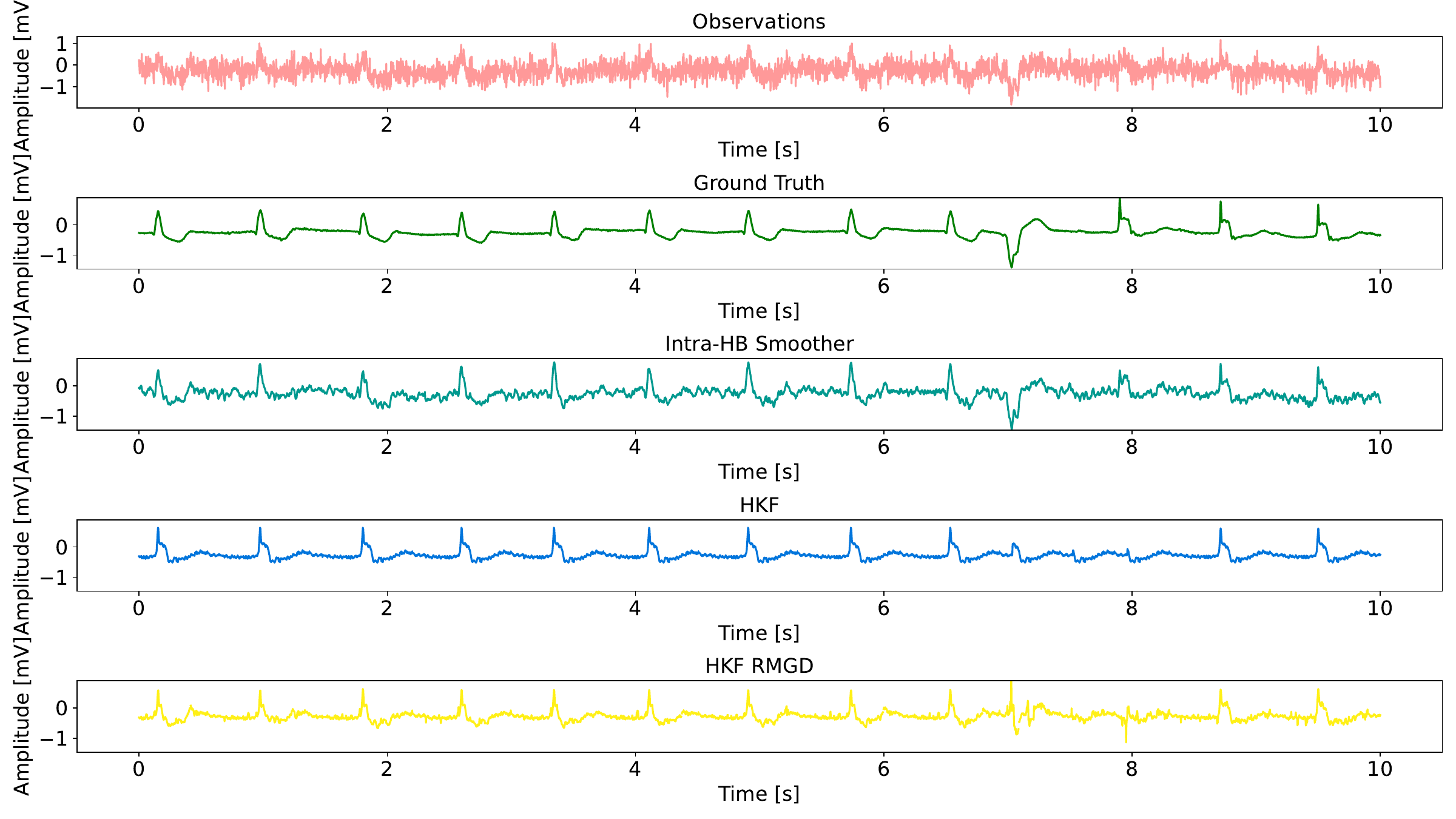}
\caption{Patient 102 - MIT-BIH}
\label{fig:mit_pat_2}  
\end{subfigure}
%
%
\begin{subfigure}[a]{0.49\columnwidth}
\centering
\includegraphics[width = 1\columnwidth]{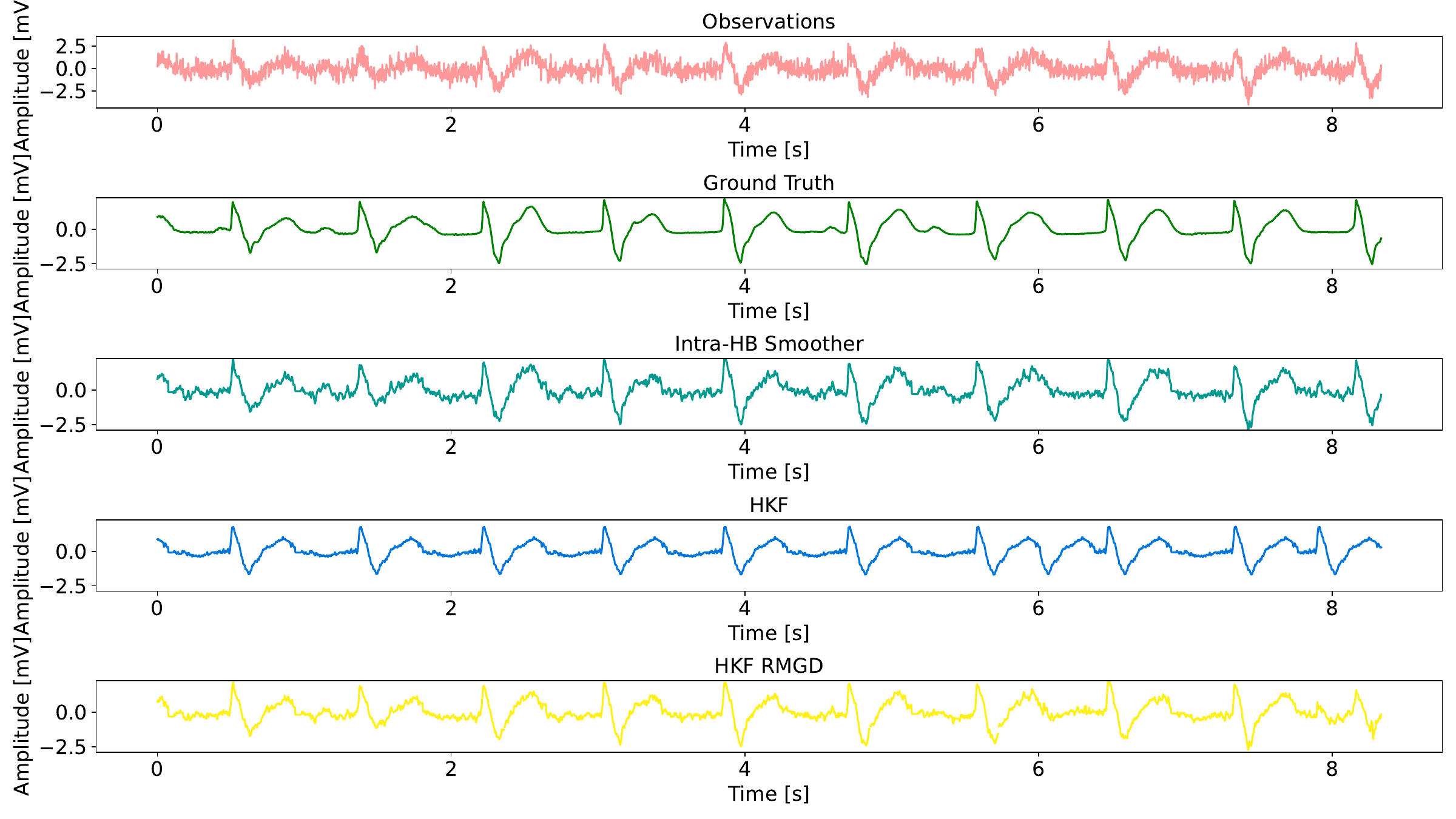}
\caption{Patient 107 - MIT-BIH}
\label{fig:mit_pat_7}  
\end{subfigure}
\caption{Sample of Consecutive \acp{hb}}
\end{figure*}

The results outlined in~\tbref{tab:mit_bih_num}\footnote{In the tables, the colors green \textcolor{green}{$\bullet$}, blue \textcolor{cyan}{$\bullet$}, and yellow \textcolor{yellow}{$\bullet$} represent rankings of 1, 2, and 3, respectively.}, showcase \ac{hkf}'s strong performance, especially when compared with alternative methods, on \ac{ecg} signals corrupted to $3\dB$ \ac{snr}. Given that the \ac{ecg} signal was acquired using $m=2$ observation channels, our \ac{hkf}, which employs a diagonal process covariance matrix, typically outperforms the \ac{rmgd} version—though performance may vary depending on the specific patient. Visual representations of our algorithm's efficacy can be found in Figures: \ref{fig:mit_pat_0_single}, \ref{fig:mit_pat_0}, \ref{fig:mit_pat_1_single}, and \ref{fig:mit_pat_1}.

The algorithm demonstrates significant capability in denoising \ac{ecg} signals. Particularly in cases free of arrhythmia, noticeable \ac{snr} enhancements are evident. It's also worth highlighting that the inter-channel filter doesn't merely force a prior onto a signal. This is apparent in~\figref{fig:mit_pat_2}, where the signal's shape shifts between the $6$th and $8$th seconds, a change that the \ac{hkf} successfully captures. This adaptability positions \ac{hkf} as an excellent tool for monitoring variations in signal shape, potentially useful in contexts like detecting heart attacks or acute hypoxia during labor.

%
%
\subsection{Patients with Arrhythmia}
Next, we delve deeper into the \ac{ecg} recordings of patients from the \mitds \acl{ds} who exhibit various types of arrhythmia, specifically, patients 102 (\figref{fig:mit_pat_2}), 107 (\figref{fig:mit_pat_7}), and 108. Arrhythmias manifest in diverse ways, ranging from accelerated heart rates (atrial fibrillation) or reduced rates (bradycardia) to sporadic episodes of unusually rapid heart rates when at rest (supraventricular tachycardia). These irregularities can be precursors to far graver conditions, including strokes~\cite{NHS_Arrhythmia}.

Given the unpredictable nature of arrhythmia in terms of shape, frequency, and amplitude, these recordings serve as an excellent benchmark for examining the upper limits of our filter. The assessment was carried out similarly to previous tests: a clean \ac{ecg} recording was corrupted by \ac{agn} to achieve a $3\dB$ \ac{snr}, and the denoised output was compared with the original version.

In situations where arrhythmia is evident, the assumption that sequential heartbeats are identical becomes invalid. Specifically, the inter-state evolution function, $\eevol{\cdot}$ from \eqref{eq:ss_inter}, doesn't remain an identity function. Consequently, from the numerical \ac{mse} findings presented in table~\ref{tab:arrhythmia_num}, it's evident that the complete \ac{hkf} is outperformed by its intermediary intra-\ac{hb} smoother output, denoted as \ac{ks}-intra. This result is logical, given that we cannot enhance performance by merging consecutive \acp{hb}. Therefore, for patients with arrhythmia, our recommendation is to solely deploy the first stage of smoothing, bypassing the subsequent stage of filtering. In such scenarios, we also advocate for harnessing the statistical modeling inherent in our algorithm to conceive an arrhythmia detection test. However, a detailed exploration of that lies beyond the purview of this paper.
%
%
\subsection{Heart-Rate Reconstruction}
\begin{figure}[h]
\centering
\includegraphics[width=0.8275\columnwidth]{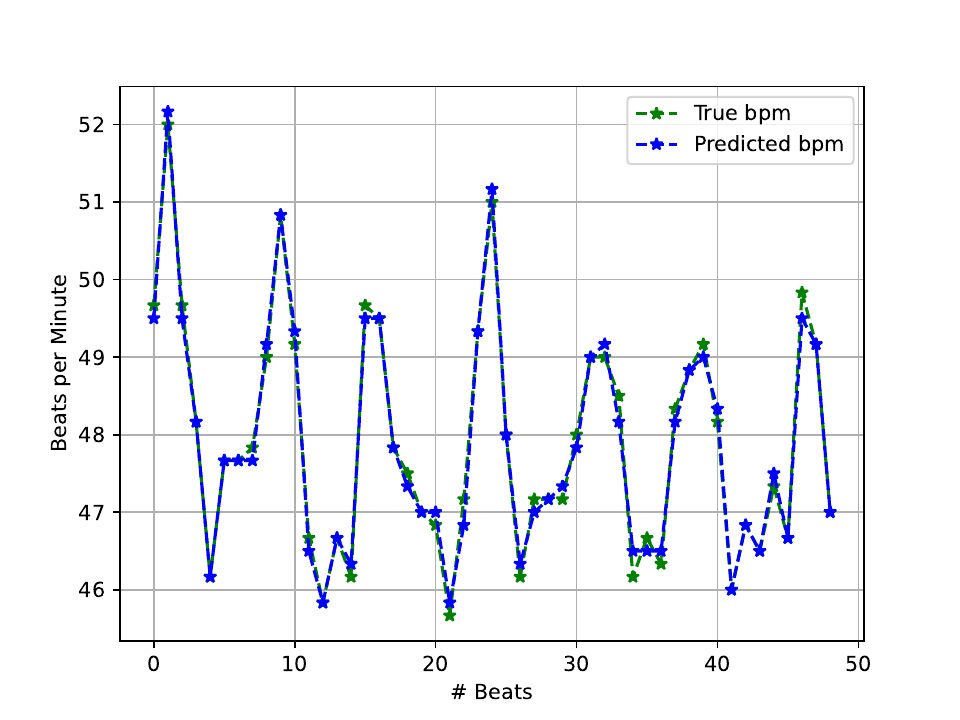}
\caption{True \ac{hr} vs. Reconstructed \ac{hr} over time}
\label{fig:mit_hr}
\end{figure}

Our algorithm's objective is to reconstruct a continuous \ac{ecg} signal from the separately filtered \acp{hb} while maintaining the integrity of the \ac{hr}. In~\figref{fig:mit_hr}, it's evident that the reconstructed \ac{hr} closely matches the true \ac{hr}, as indicated by the \acl{ds} labels.
%
%
\section{Conclusions}
In this paper, we introduced \ac{hkf}, a novel strategy for filtering noisy \ac{ecg} signals. Our approach entailed modeling the \ac{ecg} system as a hierarchical \ac{ss} system, upon which \ac{hkf} was then constructed. This design encompassed both intra- and inter-\ac{hb} dynamics, aiming to harness the maximum amount of information inherent in an \ac{ecg}. Integrating both the \ac{kf} and \ac{rts} smoothing, the method provides a clear and computationally efficient framework suitable for online filtering. Notably, the proposed method operates without the need for supervised training, relying instead on a highly patient-specific prior to adjust to variations in standard \ac{ecg} waveforms. Our results have shown that \ac{hkf} consistently surpasses other similar methods in performance. Importantly, in scenarios with significant arrhythmia in the \ac{ecg} recordings, the intermediary output of \ac{hkf} demonstrates adaptability to alterations in the inter-\ac{hb} model.
%
%
\begin{table}[!ht]
\centering
\caption{\ac{mse} $\dB$ - Proprietary Data Set - $0\dB$ \ac{snr}.}
\resizebox{\linewidth}{!}{%
\begin{tabular}{|c||c| c| c| c| c| c|} \hline \rowcolor{Gray} Patient  & Noise Floor & AE & KF-inter &  KS-intra & HKF & HKF (RMGD) \\
 \hline\hline
 0 & -17.88 & \cellcolor{white}-25.87 & \cellcolor{white}-22.66 & \cellcolor{cyan}-28.11 & \cellcolor{yellow}-27.89 & \cellcolor{green}-29.0 \\ 
 \hline 
 1 & -8.6 & \cellcolor{white}-14.42 & \cellcolor{white}-12.8 & \cellcolor{yellow}-15.37 & \cellcolor{green}-17.73 & \cellcolor{cyan}-17.34 \\ 
 \hline 
 2 & -13.66 & \cellcolor{white}-22.2 & \cellcolor{white}-19.01 & \cellcolor{yellow}-23.48 & \cellcolor{cyan}-24.8 & \cellcolor{green}-24.88 \\ 
 \hline 
 3 & -14.95 & \cellcolor{cyan}-21.91 & \cellcolor{white}-19.89 & \cellcolor{yellow}-21.53 & \cellcolor{white}-18.42 & \cellcolor{green}-22.52 \\ 
 \hline 
 4 & -16.25 & \cellcolor{white}-24.51 & \cellcolor{white}-18.75 & \cellcolor{cyan}-27.15 & \cellcolor{yellow}-25.39 & \cellcolor{green}-28.22 \\ 
 \hline 
 5 & -11.62 & \cellcolor{white}-18.36 & \cellcolor{white}-16.34 & \cellcolor{yellow}-21.64 & \cellcolor{cyan}-21.86 & \cellcolor{green}-23.57 \\ 
 \hline 
 6 & -3.94 & \cellcolor{yellow}-7.07 & \cellcolor{white}-5.74 & \cellcolor{cyan}-11.88 & \cellcolor{white}-4.93 & \cellcolor{green}-13.59 \\ 
 \hline 
 7 & -14.17 & \cellcolor{white}-19.05 & \cellcolor{white}-18.1 & \cellcolor{cyan}-23.91 & \cellcolor{yellow}-21.43 & \cellcolor{green}-25.54 \\ 
 \hline 
 8 & -16.05 & \cellcolor{yellow}-23.29 & \cellcolor{white}-19.33 & \cellcolor{cyan}-26.07 & \cellcolor{white}-19.12 & \cellcolor{green}-26.91 \\ 
 \hline 
 9 & -15.26 & \cellcolor{yellow}-20.92 & \cellcolor{white}-16.76 & \cellcolor{cyan}-24.64 & \cellcolor{white}-20.13 & \cellcolor{green}-25.87 \\ 
 \hline 
\end{tabular}
}
\label{tbl:prop}
\end{table}
%
%
\begin{table}[!ht]
\centering
\caption{\ac{mse} $\dB$ - MIT-BIH Data Set - $3\dB$ SNR}
\resizebox{\linewidth}{!}{%
\begin{tabular}{|c||c| c| c| c| c| c|} \hline \rowcolor{Gray} Patient  & Noise Floor $\dB$ & AE & KF-inter &  KS-intra & HKF & HKF (RMGD) \\
 \hline\hline
100 & -18.97 & \cellcolor{white}-20.41 & \cellcolor{white}-20.5 & \cellcolor{yellow}-25.43 & \cellcolor{green}-28.39 & \cellcolor{cyan}-25.88 \\ 
 \hline 
101 & -18.73 & \cellcolor{white}-22.31 & \cellcolor{white}-19.54 & \cellcolor{yellow}-25.92 & \cellcolor{green}-27.61 & \cellcolor{cyan}-26.25 \\ 
 \hline 
103 & -12.34 & \cellcolor{white}-16.29 & \cellcolor{white}-17.54 & \cellcolor{yellow}-20.98 & \cellcolor{green}-25.53 & \cellcolor{cyan}-22.15 \\ 
 \hline 
104 & -16.4 & \cellcolor{white}-14.72 & \cellcolor{white}-12.65 & \cellcolor{cyan}-23.76 & \cellcolor{yellow}-18.14 & \cellcolor{green}-23.96 \\ 
 \hline 
105 & -14.25 & \cellcolor{white}-20.12 & \cellcolor{white}-16.55 & \cellcolor{yellow}-23.2 & \cellcolor{cyan}-23.48 & \cellcolor{green}-23.83 \\ 
 \hline 
106 & -14.62 & \cellcolor{white}-18.96 & \cellcolor{white}-17.04 & \cellcolor{yellow}-20.68 & \cellcolor{green}-21.95 & \cellcolor{cyan}-21.52 \\ 
 \hline 
109 & -10.67 & \cellcolor{white}-15.71 & \cellcolor{white}-16.19 & \cellcolor{cyan}-20.36 & \cellcolor{yellow}-17.67 & \cellcolor{green}-21.34 \\ 
 \hline 
\end{tabular}
}
\label{tab:mit_bih_num}
\end{table}
%
%
\begin{table}[!ht]
\centering
\caption{\ac{mse} $\dB$ for the MIT-BIH Testset - 3 $\dB$ \ac{snr}.}
\resizebox{\linewidth}{!}{
\begin{tabular}{|c||c| c| c| c| c| c|} \hline \rowcolor{Gray} Patient  & Noise Floor $\dB$ & AE & KF-inter &  KS-intra & HKF & HKF (RMGD) \\
\hline\hline
102 & -4.74 & \cellcolor{white}-4.44 & \cellcolor{white}-7.98 & \cellcolor{cyan}-14.63 & \cellcolor{yellow}-12.32 & \cellcolor{green}-15.99 \\ 
\hline 
107 & -5.96 & \cellcolor{white}-6.06 & \cellcolor{yellow}-10.68 & \cellcolor{green}-14.98 & \cellcolor{white}-8.24 & \cellcolor{cyan}-12.98 \\ 
\hline 
108 & -14.27 & \cellcolor{yellow}-15.05 & \cellcolor{white}-9.8 & \cellcolor{green}-22.39 & \cellcolor{white}-12.77 & \cellcolor{cyan}-21.87 \\ 
\hline 
\end{tabular}
}
\label{tab:arrhythmia_num}
\end{table}
%

%
%
%
%
%
\begin{figure}[]
\centering
\includegraphics[width=0.52\linewidth]{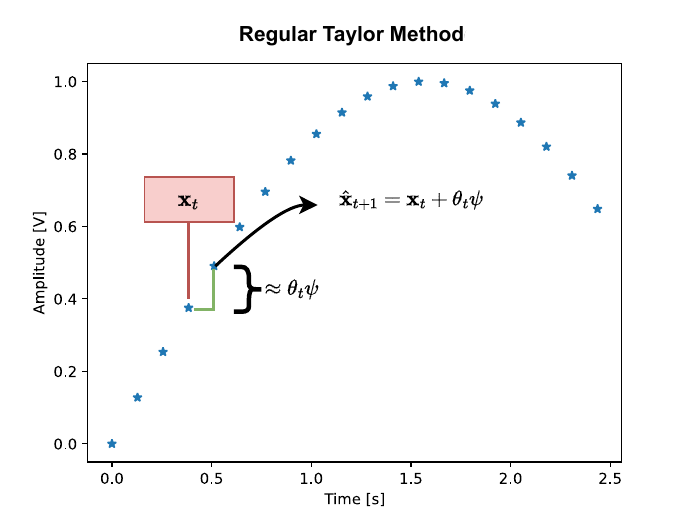}
\hspace{-0.65cm}
\includegraphics[width=0.52\linewidth]{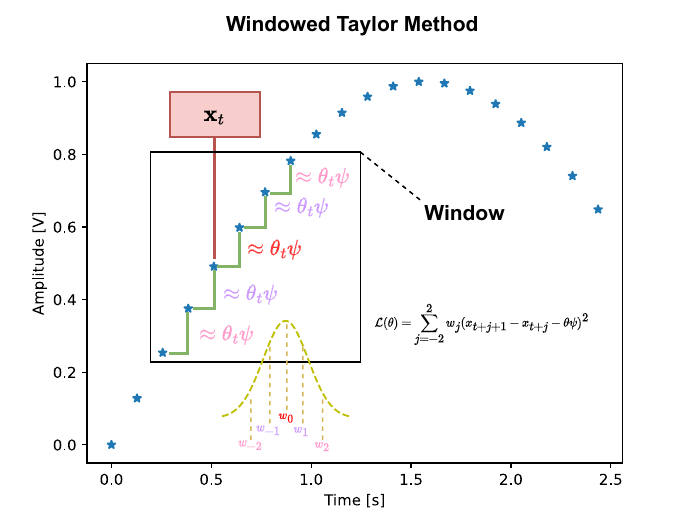}
\caption{Taylor Method}
\label{fig:taylor}
\end{figure}
%
%
%
%
%
%
%
%
%
%
%
%
%
%
%
\bibliographystyle{IEEEtran}
\bibliography{IEEEabrv,BibHKF, BibRevach, BibLoeliger}
\end{document}